\ifdefined\pdfoutput\else\newcount\pdfoutput\fi
\documentclass[a4paper,twocolumn,11pt,unpublished]{quantumarticle}
\pdfoutput=1

\usepackage[utf8]{inputenc}
\usepackage[english]{babel}
\usepackage[T1]{fontenc}
\usepackage[numbers,sort&compress]{natbib}
\usepackage{hyperref}

\usepackage{graphicx}
\usepackage{bm}
\usepackage[ruled,vlined]{algorithm2e}
\usepackage{amsmath,amssymb}
\usepackage{booktabs}
\usepackage{array}[=2016-10-06]
\usepackage{multirow}
\usepackage{xcolor}
\usepackage{comment}
\usepackage{physics}
\usepackage[caption=false]{subfig}
\usepackage{tikz}
\usetikzlibrary{arrows.meta,positioning,calc,shadows}
\usepackage{soul}
\usepackage[normalem]{ulem}

\definecolor{RED}{rgb}{1,0,0}
\definecolor{BLUE}{rgb}{0,0,1}

\begin{document}
\hypersetup{colorlinks=true,allcolors=blue}

\title{Logical Neural Belief Propagation for Linear-Complexity Decoding of Surface Codes}

\author{Hee-Youl Kwak}
\affiliation{Department of Electrical, Electronic and Computer Engineering, University of Ulsan, Ulsan 44610, South Korea}
\author{Seong-Joon Park}
\affiliation{Department of Electrical Engineering, Pohang University of Science and Technology (POSTECH), Pohang, Gyeongbuk 37673, South Korea}
\author{Dae-Young Yun}
\email{yundy@ulsan.ac.kr}
\affiliation{Department of Electrical, Electronic and Computer Engineering, University of Ulsan, Ulsan 44610, South Korea}
\author{Eliya Nachmani}
\affiliation{School of Electrical and Computer Engineering, Ben-Gurion University of the Negev, Israel}
\author{Jae-Won Kim}
\email{jaewon07.kim@gnu.ac.kr}
\affiliation{Department of Electronic Engineering, Gyeongsang National University, Jinju 52828, South Korea}

\begin{abstract}
Quantum error correction (QEC) requires accurate and efficient decoders, yet belief
propagation (BP), despite its linear decoding complexity, often provides
insufficient logical accuracy on surface codes. We propose \emph{Logical Neural
Belief Propagation} (L-NBP), a BP-based neural decoder that redirects the decoding
objective from physical-level to logical-level decoding. L-NBP uses a neural BP
(NBP) module to produce posterior beliefs, which a logical classifier transforms
into a continuous-valued soft syndrome for logical-operator prediction. Trained
end-to-end by backpropagation, the NBP module learns soft syndromes that are
favorable for logical classification. On surface codes, L-NBP matches or
outperforms BP with ordered-statistics decoding (BP-OSD) and minimum-weight perfect
matching (MWPM) while retaining the linear complexity of BP, and achieves a
threshold of $17.5\%$ under depolarizing noise. Under circuit-level noise, L-NBP
matches the accuracy of BP-OSD on the distance-$9$ surface code while requiring
only $0.2\%$ of its complexity.
\end{abstract}

\maketitle

\section{Introduction}
\label{sec_intro}

Quantum computing promises significant advantages for solving certain
classically intractable
problems~\cite{Shor1999,Grover1996,HarrowHassidimLloyd2009,AspuruGuzik2005,Peruzzo2014}.
To realize this potential at scale, quantum error correction (QEC) is
indispensable for protecting quantum information from
noise~\cite{Shor1995,Steane1996,Gottesman1996,Terhal2015}. Among existing
QEC codes, surface codes are a leading candidate owing to their high
error threshold and local
connectivity~\cite{Kitaev2003,Dennis2002,Fowler2012}. Surface codes can exhibit a
threshold behavior: once the physical error rate falls below a threshold value,
the logical error rate (LER) decreases as the code distance $d$ grows.
 
QEC decoders must achieve a low LER while operating within the coherence time of the quantum system, demanding extremely low computational complexity \cite{GoogleQuantumAI2023,GoogleQuantumAI2024, Delfosse2023}.
Minimum-weight perfect matching (MWPM)~\cite{Edmonds1965,Dennis2002,Fowler2012}
is a standard decoder for surface codes,
but it has worst-case superlinear complexity. Belief propagation (BP)~\cite{Gallager1962,MacKay2004,Poulin2008}
is attractive from a scalability viewpoint because its message-passing
complexity scales linearly with the code length. However, BP alone is not
sufficiently accurate for surface codes because of short cycles and quantum
degeneracy~\cite{Poulin2008,Babar2015}. Neural belief propagation (NBP)
improves BP by introducing trainable weights that break the symmetry of
messages~\cite{LiuPoulin2019}. Nevertheless, the accuracy of standalone BP and NBP
remains limited, and neither exhibits clear threshold behavior.

\subsection{Contributions}
\label{sec_contributions}
\begin{list}{\arabic{enumi}.}{%
  \usecounter{enumi}%
  \setlength{\leftmargin}{1.15em}%
  \setlength{\rightmargin}{0pt}%
  \setlength{\labelwidth}{0.85em}%
  \setlength{\labelsep}{0.3em}%
  \setlength{\itemsep}{0.5\baselineskip}%
  \setlength{\topsep}{0.4\baselineskip}%
  \setlength{\parsep}{0pt}%
}
\item \textbf{BP-based logical decoding.}
In this work, we propose logical NBP (L-NBP), a
BP-based neural decoder that redirects the decoding objective from
physical-level to \emph{logical-level} decoding. Rather than identifying
a physical error pattern, L-NBP couples NBP with a logical classifier that
predicts the logical operator induced by the physical error. The NBP stage
produces posterior beliefs, which the logical classifier converts into a
continuous-valued \emph{soft syndrome} and maps to the logical operator.
Because all components are trainable, L-NBP is optimized end-to-end (E2E)
with a logical-level classification loss~\cite{Varsamopoulos2017}.
The key point is that NBP is trained not to recover the underlying physical
error, but to extract a soft syndrome that is useful for logical
classification. As a result, L-NBP achieves a low LER without relying on
high-capacity architectures such as transformers~\cite{Choukroun2024,Bausch2024,Park2026};
a simple multilayer perceptron (MLP) is sufficient for logical
classification. Since both the NBP stage and the logical classifier have
linear complexity, L-NBP remains a linear-complexity decoder.

\item \textbf{Threshold improvement under depolarizing noise.}
We evaluate L-NBP under the code-capacity model with depolarizing noise.
Numerical results show that changing the decoding objective from physical
to logical substantially improves the performance over standalone NBP.
Moreover, while preserving linear-complexity scaling, L-NBP matches or
outperforms higher-complexity decoders such as
MWPM~\cite{Edmonds1965,Dennis2002,Fowler2012} and BP with ordered-statistics
decoding (BP-OSD)~\cite{PanteleevKalachev2021,RWBC2020}. Under the
code-capacity model, L-NBP attains a threshold of $17.5\%$, exceeding those
of MWPM ($14.5\%$), BP-OSD ($16.5\%$), and other high-complexity decoders
such as the quantum error correction code transformer
(QECCT)~\cite{Choukroun2024} and BP with additional memory effects
(AMBP)~\cite{KuoLai2022}.

\item \textbf{Low-complexity decoding with repeated measurements.}
We further evaluate L-NBP under the phenomenological and circuit-level noise
models~\cite{Dennis2002,Gidney2021Stim}, where syndrome measurements are
repeated over multiple rounds. Under circuit-level noise, BP-OSD operates on
the detector error model (DEM), whose graph contains many fault nodes across
measurement rounds, leading to rapidly increasing complexity. L-NBP avoids
this cost because its NBP module does not need to identify individual faults:
it only needs to extract a soft syndrome, which can be obtained from a much
more compact extended stabilizer matrix rather than from the DEM. Moreover,
because the soft-syndrome dimension is independent of the number of rounds,
the logical classifier retains the same structure as in the single-round case.
Consequently, L-NBP preserves linear $\mathcal{O}(n)$ complexity per round even
at the circuit level. At $d=9$, L-NBP achieves accuracy comparable to BP-OSD
and Relay-BP~\cite{Muller2025} while requiring only about $0.2\%$ and $0.06\%$
of their respective complexities.
\end{list}

\subsection{Related Work}
To improve BP decoding, existing approaches often add a post-processing
stage. BP-OSD~\cite{PanteleevKalachev2021,RWBC2020,Kwak2025} uses BP beliefs as
reliability information for OSD post-processing, while belief-matching~\cite{Higgott2023} converts them into MWPM edge weights. Another line of work combines neural
networks with classical decoders: neural pre-decoders preprocess the measured
syndrome or generate a residual hard syndrome before MWPM or union-find is
applied~\cite{Meinerz2022,Chamberland2026,Zhang2025}. L-NBP differs from these
approaches in that its post-processing stage is itself a simple neural network, and the NBP module and the logical classifier are
trained jointly so that they reinforce each other.

Another line of work boosts BP decoding by coupling multiple diverse BP
decoders in an ensemble-like manner, as in AMBP~\cite{KuoLai2022} and
Relay-BP~\cite{Muller2025}. In the worst case, however, these methods consume
a very large number of BP iterations---$51$ stages of $150$ iterations for AMBP
and $300$ additional stages of $60$ iterations for Relay-BP. By contrast, L-NBP fixes the
BP budget to just $60$ iterations and appends only a lightweight MLP with a
single hidden layer, which is advantageous in terms of worst-case latency.

Fully neural decoders
have also been studied for surface codes, including
feed-forward~\cite{Varsamopoulos2017,ChamberlandRonagh2018},
convolutional~\cite{JungAliHa2024}, and
transformer-based~\cite{Choukroun2024,Bausch2024,Park2026} models,
which directly learn a mapping from the measured syndrome to physical errors or
logical operators. Simple feed-forward networks are attractive from an
implementation viewpoint but offer limited
accuracy~\cite{Varsamopoulos2017,ChamberlandRonagh2018}, whereas high-capacity
architectures such as CNNs and transformers achieve stronger performance at the
cost of scalability and training
effort~\cite{JungAliHa2024,Bausch2024,Park2026}. In contrast, L-NBP relies on linear-complexity BP as its main
computational module and requires only a lightweight logical classifier,
achieving competitive logical accuracy at much lower complexity.

\subsection{Organization of the Paper}
The rest of this paper is organized as follows.
Section~\ref{sec_pre} reviews the necessary background.
Section~\ref{sec_proposed} presents the proposed L-NBP decoder: the logical classifier, the E2E training objective,
and the extension to the multi-round measurement models.
Section~\ref{sec_experiments} evaluates L-NBP in terms of both LER and
complexity, and Section~\ref{sec_conclusion} concludes the paper.

\section{Preliminaries}
\label{sec_pre}

\begin{figure}[!t]
\centering
\subfloat[]{\includegraphics[scale=0.35]{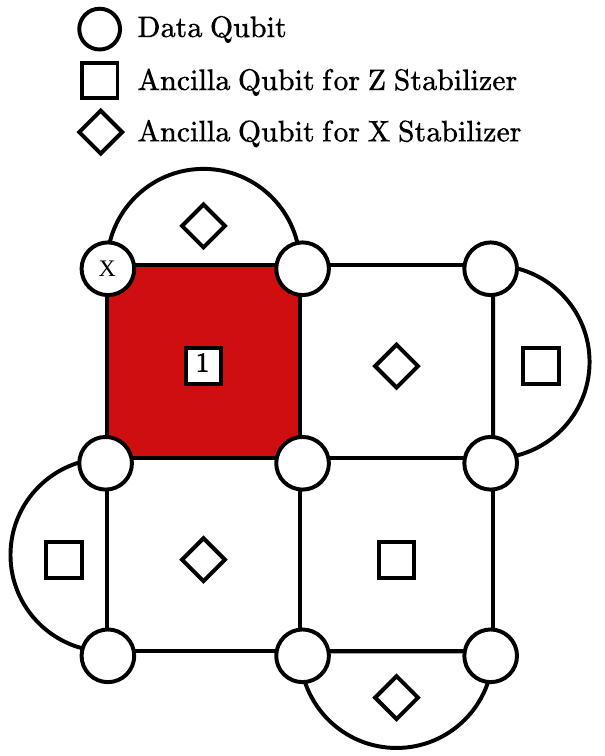}}
\hspace{20pt}
\subfloat[]{\includegraphics[scale=0.60]{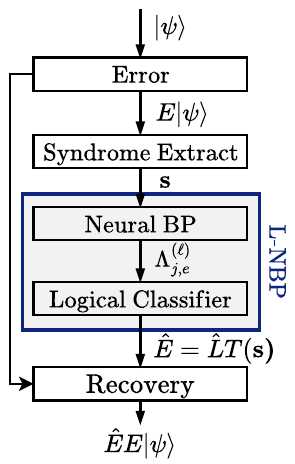}}
\caption{(a) A $d=3$ surface-code lattice with data and ancilla qubits and
stabilizers. (b) QEC workflow: a physical error $E$ induces the syndrome
$\mathbf{s}$, and L-NBP uses $\mathbf{s}$ to infer a recovery
operation.}
\label{fig:qec_workflow}
\end{figure}

\subsection{Quantum Stabilizer Codes}
An $[[n, k, d]]$ quantum stabilizer code encodes $k$ logical qubits into $n$
entangled physical qubits to protect against errors~\cite{Gottesman1996,Terhal2015}. The stabilizer group
$\mathcal{S}$ is an abelian subgroup of the $n$-qubit Pauli group
$\mathcal{P}^{\otimes n}$ with $\mathcal{P} = \{I, X, Y, Z\}$, generated by
$m = n - k$ independent, mutually commuting generators $\{S_i\}_{i=1}^{m}$. They are represented in a stabilizer matrix
$\mathbf{S} \in \{I, X, Y, Z\}^{m \times n}$. The codespace $\mathcal{C}$ is
the common $+1$ eigenspace of all stabilizers,
\begin{equation}
    \mathcal{C} = \bigl\{ \ket{\psi} \mid S\ket{\psi} = +\ket{\psi},\
    \forall S \in \mathcal{S} \bigr\},
\end{equation}
forming a $2^k$-dimensional subspace of the full $2^n$-dimensional Hilbert
space. The logical Pauli group is given by
$\mathcal{L} = N(\mathcal{S}) / \mathcal{S}$, where $N(\mathcal{S})$
denotes the set of Pauli operators that commute with all elements of
$\mathcal{S}$. The code distance $d$ is the minimum weight of any
nontrivial logical operator, i.e., any operator in
$N(\mathcal{S}) \setminus \mathcal{S}$.

Rotated surface codes, parameterized as $[[d^2, 1, d]]$, are among the most
prominent stabilizer codes because they require only local connectivity
between physical qubits \cite{Fowler2012}. As illustrated in Fig.~\ref{fig:qec_workflow}(a) for
distance $d=3$, $n = d^2$
physical (data) qubits are arranged on a two-dimensional lattice to encode a
single logical qubit ($k=1$), with $m=n-1$ stabilizers split equally into
$Z$-type (detecting $X$ errors) and $X$-type (detecting $Z$ errors). The logical operators of the surface code are represented by chains of
Pauli operators connecting opposite boundaries of the lattice. A
representative of $\bar{X}$ is a vertical chain of $d$ consecutive $X$
operators, while a representative of $\bar{Z}$ is a horizontal chain of
$d$ consecutive $Z$ operators, with $\bar{Y} = i\bar{X}\bar{Z}$.
Together with the trivial logical operator $\bar{I}$, they form the
logical group
$\mathcal{L} = \{\bar{I}, \bar{X}, \bar{Z}, \bar{Y}\}$.

A physical error $E \in \mathcal{P}^{\otimes n}$ acting on the data qubits
transforms the logical state to $E\ket{\psi}$ and is detected via the
symplectic inner product $\langle E, S_i \rangle$, which equals $0$ if $E$
and $S_i$ commute and $1$ if they anticommute. We collect these outcomes into
the binary measured syndrome
\begin{equation}
    \mathbf{s} = \mathbf{S}\odot E
    = \bigl(\langle E, S_1\rangle,\ldots,\langle E, S_m\rangle\bigr)
    \in \{0,1\}^m,
    \label{Eq:original_syndrome}
\end{equation}
where $\mathbf{S}\odot E$ denotes the syndrome map that evaluates the
symplectic product of each stabilizer generator with $E$. A stabilizer with
syndrome bit $s_i = 1$ is said to be activated by the physical error,
and inactive otherwise. Each $s_i$ is
obtained by measuring a dedicated ancilla qubit at the end of a
syndrome-extraction circuit. For example, in
Fig.~\ref{fig:qec_workflow}(a), an $X$ error on the top-left data qubit
activates its adjacent $Z$-stabilizer, yielding $s_i = 1$ on the corresponding
ancilla qubit.

\subsection{Quantum BP and NBP Decoders}
\label{subsec_bp}
In the code-capacity model, where stabilizer measurements are error-free, the
decoder produces an estimate $\hat{E}$ from the measured syndrome $\mathbf{s}$
of size $m$.
Quantum BP decoding estimates physical errors by iteratively passing
log-likelihood ratio (LLR) messages over a Tanner graph, which consists of
$n$ variable nodes (VNs) corresponding to physical qubits and $m$ check nodes
(CNs) corresponding to stabilizers. 
Each CN $c_i$ is initialized with the corresponding measured syndrome bit
$s_i$, and the decoder exchanges messages between CNs and VNs over a maximum of
$\overline{\ell}$ iterations in order to infer whether an error has occurred on
each qubit. At every iteration $\ell \in \{1,\dots,\overline{\ell}\}$, the decoder produces a posterior LLR
$\Lambda_{j,e}^{(\ell)}$ for each VN $v_j$ and error type
$e\in\{X,Y,Z\}$, which quantifies the belief that $E_j = I$ relative to
$E_j = e$. The complete message update rules are deferred to
Appendix~\ref{app_bp}. The estimated error $\hat{E}_j$ for VN $v_j$ is read
from the posterior LLRs,
\begin{equation}
    \hat{E}_j = \begin{cases}
    I & \text{if } \Lambda_{j,e}^{(\overline{\ell})} > 0 \text{ for all } e, \\
    \mathop{\mathrm{arg\,min}}\limits_{e}~\Lambda_{j,e}^{(\overline{\ell})} & \text{otherwise},
    \end{cases}
    \label{Eq:Hard_Decision}
\end{equation}
and the decoded syndrome $\hat{\mathbf{s}} = \mathbf{S}\odot\hat{E}$ is
computed from $\hat{E}$. The decoding outcome falls into one of three cases:
\begin{enumerate}
    \item If $\hat{\mathbf{s}} \neq \mathbf{s}$, a flagged failure is declared.
    \item If $\hat{\mathbf{s}} = \mathbf{s}$ but $E\hat{E} \in \mathcal{L}$, an
    unflagged failure is declared and the logical state is changed.
    \item If $\hat{\mathbf{s}} = \mathbf{s}$ and $E\hat{E} \in \mathcal{S}$,
    decoding succeeds and the logical state is preserved.
\end{enumerate}
The LER is then the sum of the flagged and unflagged failure probabilities.
While unflagged failures are undetectable, flagged failures can be handled by
post-processing. In particular, the posterior LLRs $\Lambda_{j,e}^{(\ell)}$ can
be forwarded to a post-processing stage.

NBP retains this message-passing structure but replaces the plain sums in the
VN and posterior updates with weighted sums, whose trainable weights are
collected in a parameter set $\theta_{\mathrm{N}}$ defined in
\eqref{Eq:NBP_Params}. In classical BP, all of these weights are fixed to one,
whereas weighted BP applies a single fixed scaling factor to the CN messages.
NBP with parameters $\theta_{\mathrm{N}}$ generates $\hat{E}$ and is trained to
estimate the physical error pattern $E$ using the loss function
\begin{equation}
    \mathsf{L}_{\mathrm{NBP}}(\theta_{\mathrm{N}}) =
    \Pr\!\left[ \hat{E}\, E \notin \mathcal{S} \right],
    \label{eq:loss_nbp}
\end{equation}
which coincides with the LER; a differentiable
surrogate~\cite{LiuPoulin2019, Miao2025} is used in practice for training.

\begin{figure*}[!t]
\centering
\includegraphics[scale=0.4]{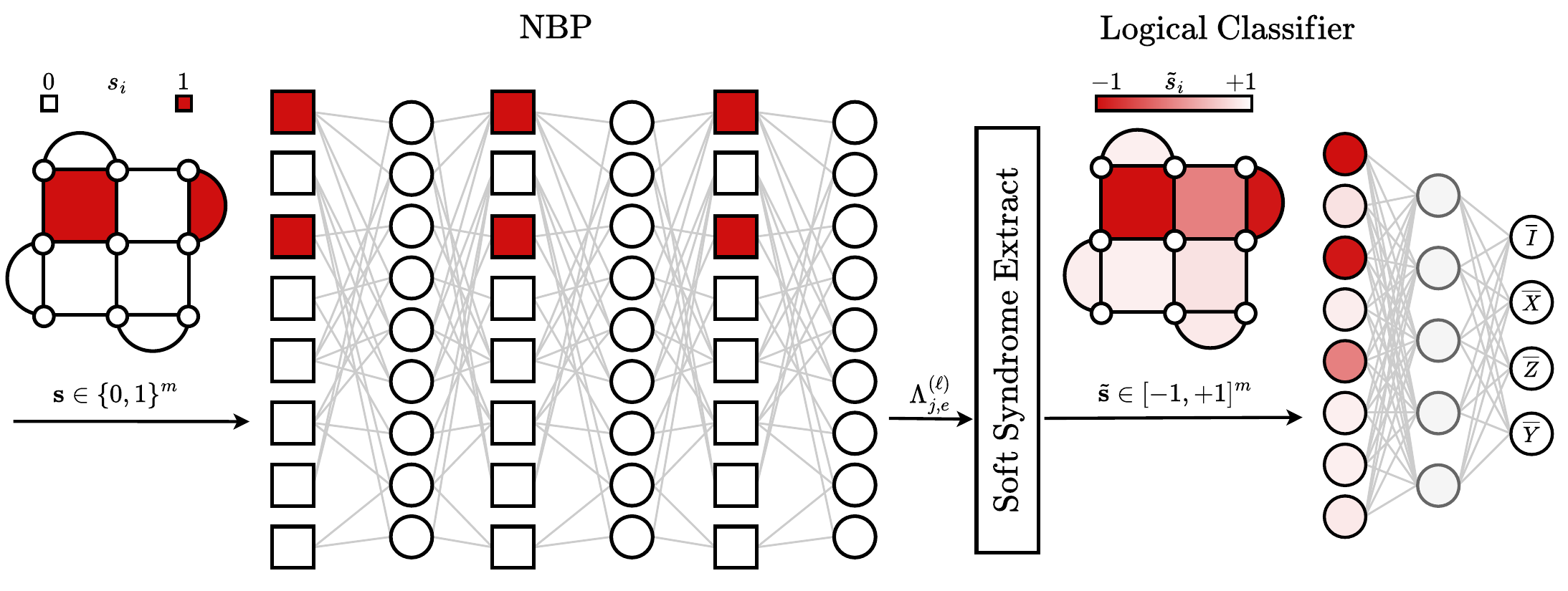}
\caption{The proposed L-NBP decoder consists of two stages. The NBP module
processes the measured hard syndrome $\mathbf{s}$ and produces posterior
beliefs ${\Lambda}_{j,e}^{(\ell)}$; the logical classifier then converts these beliefs into a soft
syndrome $\tilde{\mathbf{s}}$ and maps it to the logical operator. All stages of
L-NBP are trained end-to-end for accurate logical classification.}
\label{fig:lnbp_architecture}
\end{figure*}

\subsection{Logical-Level Decoding}
\label{subsec_logical_class_decoding}
The BP decoder above is a physical-level decoder that estimates the error on each
individual physical qubit. For QEC, however, the exact physical error is not
the final object of interest. Decoding can therefore be formulated as
logical-level decoding~\cite{Varsamopoulos2017,JungAliHa2024}.
For a stabilizer code, any physical error can be decomposed as
\begin{equation}
    E = T(\mathbf{s})\, L\, S,
    \label{Eq:Error_Decomposition}
\end{equation}
where $T(\mathbf{s}) \in \mathcal{P}^{\otimes n}$ is a pure error, $L \in \mathcal{L}$ is a logical operator, and $S \in \mathcal{S}$ is a stabilizer element. The pure
error is a low-weight Pauli operator that reproduces the measured syndrome
$\mathbf{s}$ and is constructed from precomputed elementary pure errors associated with the syndrome bits.
Applying the recovery $T(\mathbf{s})$ maps the
corrupted state $E\ket{\psi}$ back into the codespace,
\begin{equation}
    T(\mathbf{s})\, E \ket{\psi} = L\, S \ket{\psi} = L\ket{\psi}.
\end{equation}
Thus, the remaining decoding task is to estimate the logical operator $L$.

For the surface code with one logical qubit, the logical operators are
$\{\bar{I},\bar{X},\bar{Z},\bar{Y}\}$. A logical-level decoder produces an
estimated logical operator $\hat{L}$ and applies the recovery
$\hat{E} = \hat{L}T(\mathbf{s})$. If $\hat{L} = L$, then
$\hat{E}E\ket{\psi} = S\ket{\psi} = \ket{\psi}$ and decoding succeeds; otherwise, the
residual logical operator $\hat{L}L \neq \bar{I}$ corrupts the logical state and
results in a logical failure.

Although both physical- and logical-level decoders ultimately output a recovery
$\hat{E}$, they differ in what they estimate: a physical-level decoder infers
$\hat{E}$ directly, whereas a logical-level decoder only estimates the logical
operator $\hat{L}$ and obtains the recovery deterministically as
$\hat{E} = \hat{L}\,T(\mathbf{s})$, since the pure error $T(\mathbf{s})$ is fixed
by the measured syndrome. The main task therefore reduces to a four-way
classification over $\{\bar{I},\bar{X},\bar{Z},\bar{Y}\}$.

\section{Logical Neural Belief Propagation}
\label{sec_proposed}
The overall error-correction process for the proposed L-NBP decoder is
illustrated in Fig.~\ref{fig:qec_workflow}(b), with its detailed architecture
shown in Fig.~\ref{fig:lnbp_architecture}. L-NBP first runs the NBP module on the measured
syndrome $\mathbf{s}$ and appends a logical classifier that converts the
resulting posterior LLRs into a soft syndrome and maps it to the logical
operator $\hat{L}$, from which the final recovery $\hat{E}=\hat{L}\,T(\mathbf{s})$
is obtained.

\subsection{Logical Classifier}
\label{subsec_logical_classifier}
For NBP, the decoded syndrome $\hat{\mathbf{s}} = \mathbf{S}\odot\hat{E}$ is obtained from the estimated error $\hat{E}$. Each bit $\hat{s}_i$ is
hard-valued, equal to $0$ if CN $i$ is inactive and $1$ if it is active. The NBP
module, however, provides soft information about the estimated error, so we can
replace the hard decoded syndrome with a \emph{soft syndrome} whose value
reflects the reliability that each CN is activated.
From the posterior LLRs $\Lambda_{j,e}^{(\ell)}$ in \eqref{Eq:Post_Sum}
produced by the NBP module, we
compute an edge-wise posterior LLR
$\Lambda_{j\to i}^{(\ell)}$ that represents the belief that the estimated
error on qubit $j$ commutes or anticommutes with stabilizer $S_i$. Applying
the same derivation as in \eqref{Eq:Belief_Quantization}, for a $Z$-type
stabilizer,
\begin{equation}
    \Lambda_{j \to i}^{(\ell)}
    = \zeta(-\Lambda_{j,Z}^{(\ell)}) + \Lambda_{j,X}^{(\ell)}
      - \zeta(\Lambda_{j,X}^{(\ell)} - \Lambda_{j,Y}^{(\ell)}),
    \label{Eq:Marginal_LLR_Z}
\end{equation}
and for an $X$-type stabilizer,
\begin{equation}
    \Lambda_{j \to i}^{(\ell)}
    = \zeta(-\Lambda_{j,X}^{(\ell)}) + \Lambda_{j,Z}^{(\ell)}
      - \zeta(\Lambda_{j,Z}^{(\ell)} - \Lambda_{j,Y}^{(\ell)}),
    \label{Eq:Marginal_LLR_X}
\end{equation}
where $\zeta(x) = \ln(1+e^{x})$ is the softplus function.
These edge-wise posterior LLRs are aggregated at each CN via the min-sum rule,
yielding the soft syndrome at CN $i$ and iteration $\ell$:
\begin{equation}
    \overline{s}_i^{(\ell)} = \left( \prod_{j \in \mathcal{N}_c(i)}
    \mathrm{sgn}(\Lambda_{j \to i}^{(\ell)}) \right)
    \min_{j \in \mathcal{N}_c(i)} \left|\Lambda_{j \to i}^{(\ell)}\right|.
    \label{Eq:Per_Iter_Soft}
\end{equation}

To extract richer soft information from the NBP module, the soft syndromes
across iterations are combined through a learnable weighted
sum,
\begin{equation}
    \overline{s}_i = \sum_{k=1}^{K}
    \gamma^{(k)}\, \overline{s}_i^{(k\Delta)},
    \qquad K = \overline{\ell}/\Delta,
    \label{Eq:Soft_Syndrome_Agg}
\end{equation}
where $\Delta$ is the iteration sampling interval, $K = \overline{\ell}/\Delta$
is the number of aggregated soft syndromes, and $\{\gamma^{(k)}\}_{k=1}^{K}$
are trainable weights summing to one. Finally,
$\overline{s}_i$ is normalized by a learnable temperature $\tau$ and passed
through a $\tanh$ nonlinearity:
\begin{equation}
    \tilde{s}_i = \tanh\!\left( \frac{(-1)^{s_i}\, \overline{s}_i}{\tau} \right).
    \label{Eq:soft_syndrome}
\end{equation}
The factor $(-1)^{s_i}$ incorporates the measured syndrome bit $s_i$, so that
$\tilde{s}_i$ is a \emph{residual} soft syndrome that combines the NBP belief
with the measured syndrome.
The resulting soft syndrome $\tilde{s}_i$ takes values in $[-1,1]$, while its magnitude reflects the confidence of the soft decision.

Next, the MLP takes the soft syndrome $\tilde{\mathbf{s}} \in [-1,1]^m$ as
input and produces a four-dimensional logit vector $\mathbf{z}$, one entry per
logical operator $\{\bar{I},\bar{X},\bar{Z},\bar{Y}\}$. It is realized with a
single hidden layer,
\begin{align}
    \mathbf{h} &= \tanh(W_h\,\tilde{\mathbf{s}} + \mathbf{b}_h),\\
    \mathbf{z} &= W_o\,\mathbf{h} + \mathbf{b}_o
    \in \mathbb{R}^4,
\end{align}
where $W_h \in \mathbb{R}^{h \times m}$, $\mathbf{b}_h \in \mathbb{R}^{h}$,
$W_o \in \mathbb{R}^{4 \times h}$, $\mathbf{b}_o \in \mathbb{R}^{4}$, and $h$ is
the hidden dimension. The predicted logical operator is
$\hat{L} = \arg\max_c z_c$, and the final recovery operator is
$\hat{E}=\hat{L}\,T(\mathbf{s})$. As in
Section~\ref{subsec_logical_class_decoding}, decoding succeeds if $\hat{L}$
matches the true logical operator and fails otherwise.

\subsection{L-NBP Decoding}
\label{subsec_e2e_training}
Let $f^{\mathrm{C}}_{\theta_{\mathrm{C}}}$ denote the logical classifier just
described, which maps the posterior LLRs $\Lambda_{j,e}^{(\ell)}$ to the logit
vector $\mathbf{z}$, with trainable parameters
$\theta_{\mathrm{C}} = \{\gamma^{(k)},\tau,W_h,\mathbf{b}_h,W_o,\mathbf{b}_o\}$.
Combining the NBP and classifier stages, the complete L-NBP decoder is the composition
\begin{equation}
    f^{\mathrm{L}}_{\Theta}
    = f^{\mathrm{C}}_{\theta_{\mathrm{C}}} \circ
      f^{\mathrm{N}}_{\theta_{\mathrm{N}}},
    \qquad
    \Theta = \{\theta_{\mathrm{N}},\theta_{\mathrm{C}}\},
    \label{Eq:LNBP_Composition}
\end{equation}
which maps a measured hard syndrome $\mathbf{s}$ to a logit vector
$f^{\mathrm{L}}_{\Theta}(\mathbf{s}) \in \mathbb{R}^4$ and predicts
$\hat{L}=\arg\max_c\,[f^{\mathrm{L}}_{\Theta}(\mathbf{s})]_c$. The two stages act
in sequence, as summarized in
Algorithm~\ref{alg:lnbp}: the NBP module $f^{\mathrm{N}}_{\theta_{\mathrm{N}}}$
runs trainable message passing on $\mathbf{s}$ and produces the posterior LLRs
$\Lambda_{j,e}^{(\ell)}$, which the logical classifier
$f^{\mathrm{C}}_{\theta_{\mathrm{C}}}$ then converts into the soft syndrome
$\tilde{\mathbf{s}}$ and maps to the output logit vector.

\begin{algorithm}[t]
\caption{L-NBP decoding}
\label{alg:lnbp}
\KwIn{Measured hard syndrome $\mathbf{s}$}
\tcc{Stage 1: NBP message passing}
Initialize $\mu_{j\to i,e}^{(0)} = \Lambda_{j,e}^{(0)}$ from the initial prior LLR\;
\For{$\ell=1$ \KwTo $\overline{\ell}$}{
    Belief quantization: compute $\lambda_{j\to i}^{(\ell-1)}$ via \eqref{Eq:Belief_Quantization}\;
    CN update: compute $\nu_{i\to j}^{(\ell)}$ via \eqref{Eq:BP_CtV}\;
    VN update: compute $\mu_{j\to i,e}^{(\ell)}$ via \eqref{Eq:Weighted_Sum}--\eqref{Eq:Damping}\;
    Decision: compute posterior LLR $\Lambda_{j,e}^{(\ell)}$ via \eqref{Eq:Post_Sum}\;
}
\tcc{Stage 2: Logical classification}
\For{$k=1$ \KwTo $K$ \textnormal{(}$K=\overline{\ell}/\Delta$\textnormal{)}}{
    Set $\ell = k\Delta$\;
    Compute $\Lambda_{j\to i}^{(\ell)}$ via \eqref{Eq:Marginal_LLR_Z}--\eqref{Eq:Marginal_LLR_X}\;
    Compute $\overline{s}_i^{(\ell)} = \bigl(\prod_{j}\mathrm{sgn}(\Lambda_{j\to i}^{(\ell)})\bigr)
    \min_{j}|\Lambda_{j\to i}^{(\ell)}|$\;
}
Aggregate: $\overline{s}_i = \sum_{k=1}^{K}\gamma^{(k)}\overline{s}_i^{(k\Delta)}$\;
Normalize: $\tilde{s}_i = \tanh\bigl((-1)^{s_i}\overline{s}_i/\tau\bigr)$\;
MLP: $\mathbf{h} = \tanh(W_h\tilde{\mathbf{s}} + \mathbf{b}_h)$,\quad
$\mathbf{z} = W_o\mathbf{h} + \mathbf{b}_o$\;
\KwOut{Predicted logical operator $\hat{L}=\arg\max_c\, z_c$}
\end{algorithm}

\begin{figure}[!t]
\centering
\includegraphics[scale=0.4]{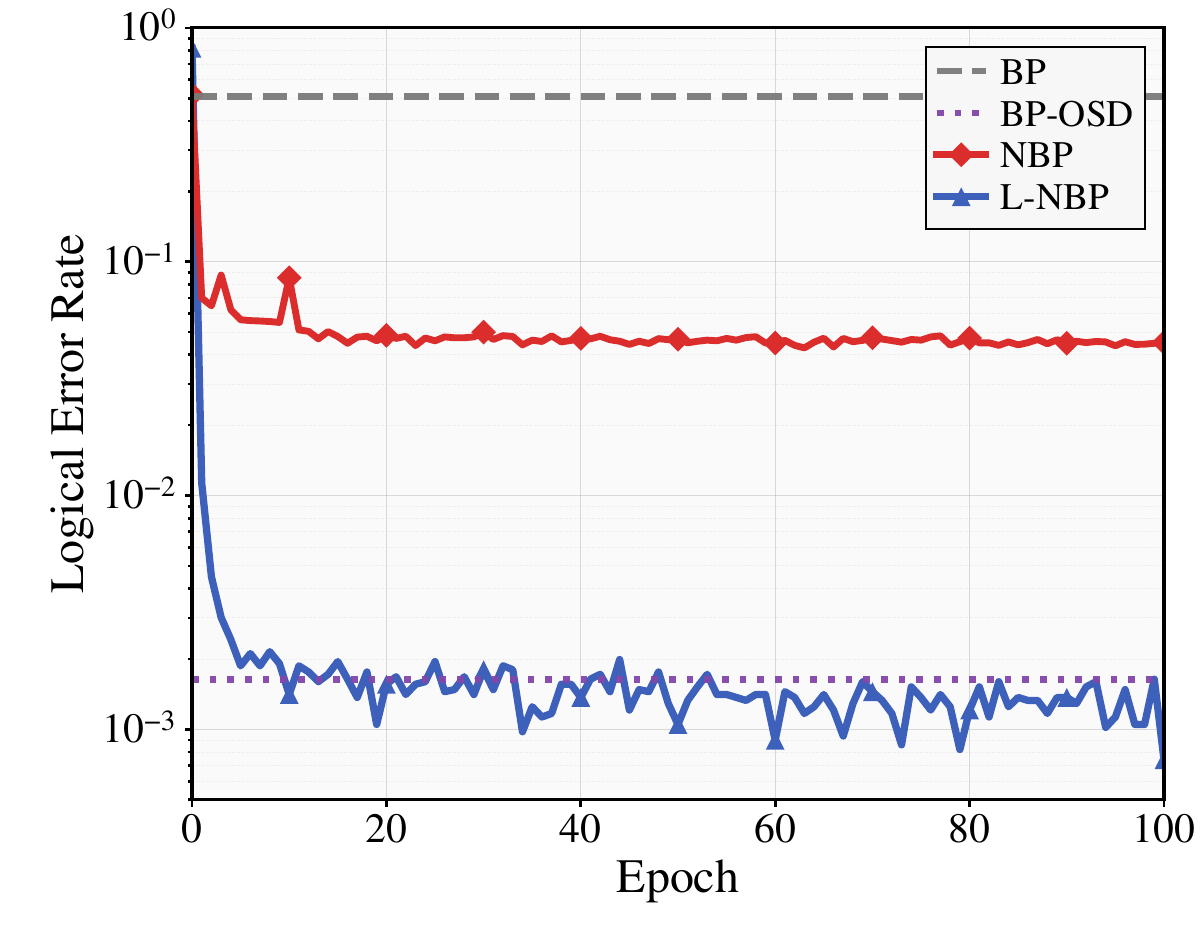}
    \caption{LER versus training epochs for the $d=9$ surface
    code under the code-capacity model. Both NBP and L-NBP improve on BP, but only
    L-NBP converges to a substantially lower LER and surpasses the BP-OSD
    baseline.}
\label{Fig:train_curve}
\end{figure}

All components of L-NBP are piecewise differentiable, so the decoder is
trainable end-to-end and is optimized with the logical-level cross-entropy loss
\begin{equation}
    \mathsf{L}_{\mathrm{L\text{-}NBP}}(\Theta)
    = \mathrm{CE}\!\left(
        f^{\mathrm{L}}_{\Theta}(\mathbf{s}),\, L
      \right),
    \label{eq:loss_lnbp}
\end{equation}
where $L$ is the true logical operator. Gradients are backpropagated through
the logical classifier into the NBP module.
Notably, we do not additionally train the NBP module with a physical-level loss
such as \eqref{eq:loss_nbp}, which targets the NBP module's own LER. Such a loss
would bias the NBP module toward its own physical-level decoding, whereas L-NBP
requires the NBP beliefs to be useful for the final logical-level
classification. The training hyperparameters, including the hidden dimension
$h=256$ and the number of BP iterations $\overline{\ell}=60$, are
summarized in Appendix~\ref{app_training}.

Fig.~\ref{Fig:train_curve} shows the LER at the physical error rate $p=0.05$ over training epochs for
the $d=9$ surface code under the code-capacity model. NBP improves on BP but
saturates at a limited LER. By adopting the logical-decoding objective, L-NBP
achieves a much larger improvement and surpasses the BP-OSD baseline.

\subsection{Qualitative Analysis via t-SNE Visualization}
\label{subsec_tsne}

\begin{figure*}[!t]
\centering
\subfloat[]{\includegraphics[width=0.24\textwidth]{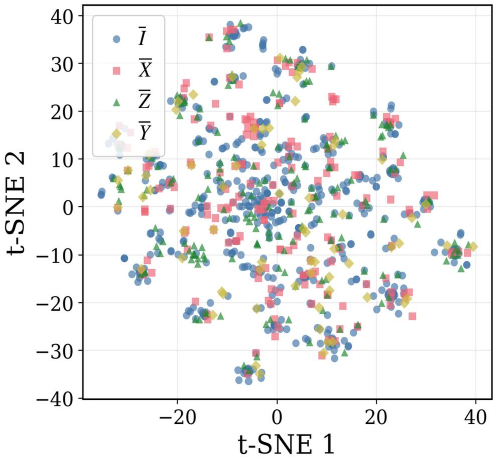}}
\subfloat[]{\includegraphics[width=0.24\textwidth]{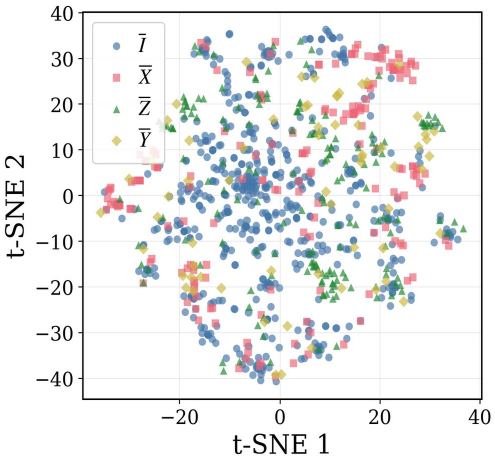}}
\subfloat[]{\includegraphics[width=0.24\textwidth]{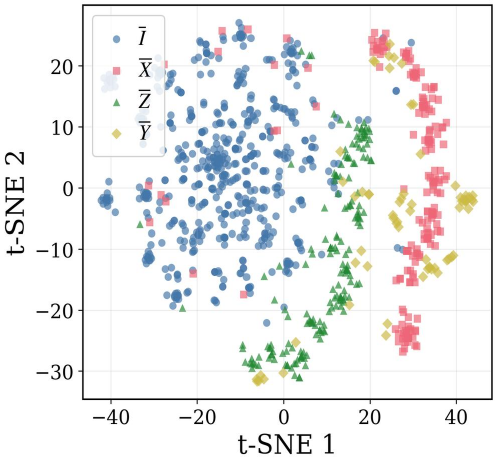}}
\subfloat[]{\includegraphics[width=0.24\textwidth]{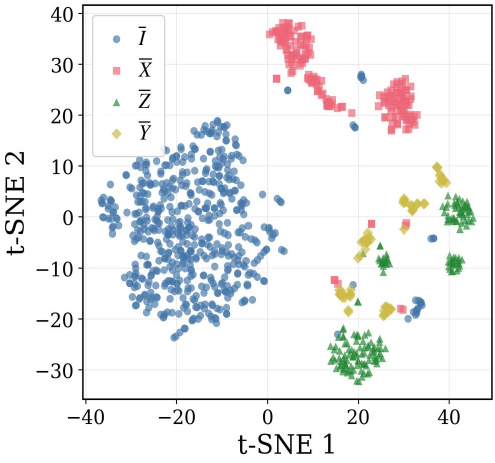}}
    \caption{t-SNE visualization of the L-NBP soft syndrome
    $\tilde{\mathbf{s}}$ at successive training stages, after (a) $100$,
    (b) $500$, and (c) $1000$ training batches, and (d) at the end of training
    (surface code $d=9$, $p=0.05$). Each color represents one of the four
    logical operators $\{\bar{I},\bar{X},\bar{Z},\bar{Y}\}$. As training
    proceeds, the soft syndrome separates into increasingly compact and
    well-separated clusters by logical operator.}
\label{Fig:t_SNE}
\end{figure*}

To understand why the soft syndrome is effective and why a simple logical
classifier suffices, we use t-SNE~\cite{VanDerMaaten2008} to visualize how the
soft syndrome $\tilde{\mathbf{s}}$ produced by the NBP stage evolves during
training. Fig.~\ref{Fig:t_SNE} shows $1000$ samples collected for the surface
code at $d=9$ and $p=0.05$, each colored by its logical operator. Early in
training after $100$ batches (Fig.~\ref{Fig:t_SNE}(a)), the four logical
operators are heavily overlapped and largely indistinguishable. As training
progresses, they begin to pull apart and form compact, clearly separated clusters as shown in Fig.~\ref{Fig:t_SNE}(d) by the end of training. In other words, the E2E objective drives the NBP stage to learn a soft-syndrome
representation in which the lightweight classifier can easily separate the logical operators.

\subsection{Extension to Multi-Round Measurement Models}
\label{subsec_pheno}
So far, we have assumed the code-capacity noise model, in which stabilizer
measurements are perfect and the measured syndrome is noiseless. We now
consider the multi-round measurement models, namely the
phenomenological~\cite{Dennis2002,KuoChernLai2021,KuoLai2025} and
circuit-level~\cite{Fowler2012,Gidney2021Stim,Higgott2023} noise models, in
which stabilizer measurements are repeated over
multiple rounds and the measured syndrome itself can be faulty.

The phenomenological noise model
performs $R$ rounds of syndrome extraction,
typically set to $R = d$ for surface codes. In each round
$r \in \{0, 1, \dots, R-1\}$, a Pauli error $F_r \in \mathcal{P}^{\otimes n}$
acts on the data qubits with probability $p/3$ per error type, and the
measurement is corrupted by a binary measurement error
$\mathbf{m}_r \in \{0,1\}^m$ with probability $p$ per bit. An additional syndrome readout round at $r = R$ is appended as a boundary
condition, where $F_R \in \mathcal{P}^{\otimes n}$ acts on the data qubits
and the syndrome measurement is error-free, i.e.,
$\mathbf{m}_R = \mathbf{0}$.
Because per-round errors accumulate on the data qubits, the decoding target of
physical-level decoders is the cumulative error $E = F_0 F_1 \cdots F_R$. The
syndrome measured at round $r$ reflects the accumulated error
$F_0 \cdots F_r$ and is corrupted by $\mathbf{m}_r$:
\begin{equation}
    \mathbf{s}_r \;=\; \mathbf{S}\odot(F_0 \cdots F_r)
    \oplus \mathbf{m}_r \;\in\; \{0,1\}^m,
\end{equation}
where $\oplus$ denotes the entrywise modulo-2 (XOR) addition.

The BP decoder operates on the detectors
$\mathbf{d}_r = \mathbf{s}_r \oplus \mathbf{s}_{r-1}$, with the convention
$\mathbf{s}_{-1} = \mathbf{0}$ \cite{Higgott2023,KuoLai2025}. Taking
$\{\mathbf{d}_r\}_{r=0}^{R}$ as input, it jointly estimates
the per-round data errors $\{F_r\}_{r=0}^{R}$ and the measurement errors
$\{\mathbf{m}_r\}_{r=0}^{R-1}$~\cite{KuoChernLai2021,KuoLai2025}. Decoding
runs on an extended stabilizer matrix $\mathbf{S}_{\mathrm{ext}}$ such as
\begin{equation}
\mathbf{S}_{\mathrm{ext}}
=\left[
\begin{array}{cccc|cccc}
\mathbf{S} & \mathbf{0} & \cdots & \mathbf{0} & \mathbf{I} & \mathbf{0} & \cdots & \mathbf{0} \\
\mathbf{0} & \mathbf{S} & \cdots & \mathbf{0} & \mathbf{I} & \mathbf{I} & \ddots & \vdots \\
\vdots & \vdots & \ddots & \vdots & \mathbf{0} & \ddots & \ddots & \mathbf{0} \\
\mathbf{0} & \mathbf{0} & \cdots & \mathbf{S} & \mathbf{0} & \cdots & \mathbf{I} & \mathbf{I}
\end{array}
\right].
\label{Eq:Extended_S_Matrix}
\end{equation}
The extended matrix has $m(R+1)$ rows for detectors, $n(R+1)$ columns for
data errors, and $mR$ columns for measurement errors.  The cumulative error estimate is
$\hat{E} = \hat{F}_0\cdots \hat{F}_R$. In the NBP variant, the message-passing
weights are made trainable.

For logical-level decoding, as in the code-capacity case, L-NBP predicts the
logical operator $L$ in the decomposition $E = T(\mathbf{s}_R)\,L\,S$, where the
pure error $T(\mathbf{s}_R)$ is obtained from the measured final-round syndrome
$\mathbf{s}_R$. The NBP module runs on the extended matrix
$\mathbf{S}_{\mathrm{ext}}$, takes $\{\mathbf{d}_r\}_{r=0}^{R}$ as input, and
produces per-round posterior LLRs. To obtain the soft syndrome from these LLRs,
we exploit the linearity of the syndrome map. The decoded final syndrome is
$\hat{\mathbf{s}}_R = \mathbf{S}\odot \hat{E}
= \mathbf{S}\odot(\hat{F}_0\cdots\hat{F}_R)
= \bigoplus_{r=0}^{R}\mathbf{S}\odot \hat{F}_r$, so the estimated error
$\hat{F}_r$ for each round contributes to $\hat{\mathbf{s}}_R$ through the same
stabilizer connections in $\mathbf{S}$. Since an XOR in the binary domain
corresponds to the min-sum rule in the LLR domain, the soft syndrome for each CN
can be computed as
\begin{equation}
    \begin{aligned}
    \overline{s}_i^{(\ell)}
    &= \Bigg(\prod_{r=0}^{R}\prod_{j \in \mathcal{N}_c(i)}
      \mathrm{sgn}\bigl(\Lambda_{(j,r)\to i}^{(\ell)}\bigr)\Bigg)
      \\
      &\quad{}\times \min_{\substack{0 \le r \le R \\ j \in \mathcal{N}_c(i)}}
      \bigl|\Lambda_{(j,r)\to i}^{(\ell)}\bigr|,
    \label{Eq:MS_Round}
    \end{aligned}
\end{equation}
for $i = 1,\dots,m$, where $\Lambda_{(j,r)\to i}^{(\ell)}$ is the edge-wise
posterior LLR sent from the VN at qubit $j$ in round $r$ to CN $i$ at iteration
$\ell$. Aggregating over iterations and combining with the measured syndrome
$\mathbf{s}_R$ as in
\eqref{Eq:Soft_Syndrome_Agg}--\eqref{Eq:soft_syndrome} then yields a soft
syndrome $\tilde{\mathbf{s}} \in [-1,1]^m$.

The resulting soft syndrome has length $m = n-1$ regardless of
$R$; that is, the NBP module absorbs the syndrome history
$\{\mathbf{s}_r\}_{r=0}^{R}$ across the temporal dimension and acts as a
syndrome-history compressor. Consequently, the logical classifier needs no
architectural change between single-round and multi-round decoding, and its
input size stays $m$.

Under the circuit-level noise model, we perform a $Z$-memory experiment using the Stim stabilizer-circuit
simulator~\cite{Gidney2021Stim}. The Stim simulator provides a DEM, which is converted into a binary
parity-check matrix $\mathbf{H}_{\mathrm{DEM}}$ of size $N_d \times N_f$, whose $N_d$
rows correspond to detectors and whose $N_f$ columns correspond to faults. BP-OSD
runs BP on $\mathbf{H}_{\mathrm{DEM}}$ to obtain posterior LLRs for the faults,
which are then post-processed by OSD.

\begin{figure}[!t]
\centering
\subfloat[$d=13$, code-capacity]{\includegraphics[width=0.88\columnwidth]{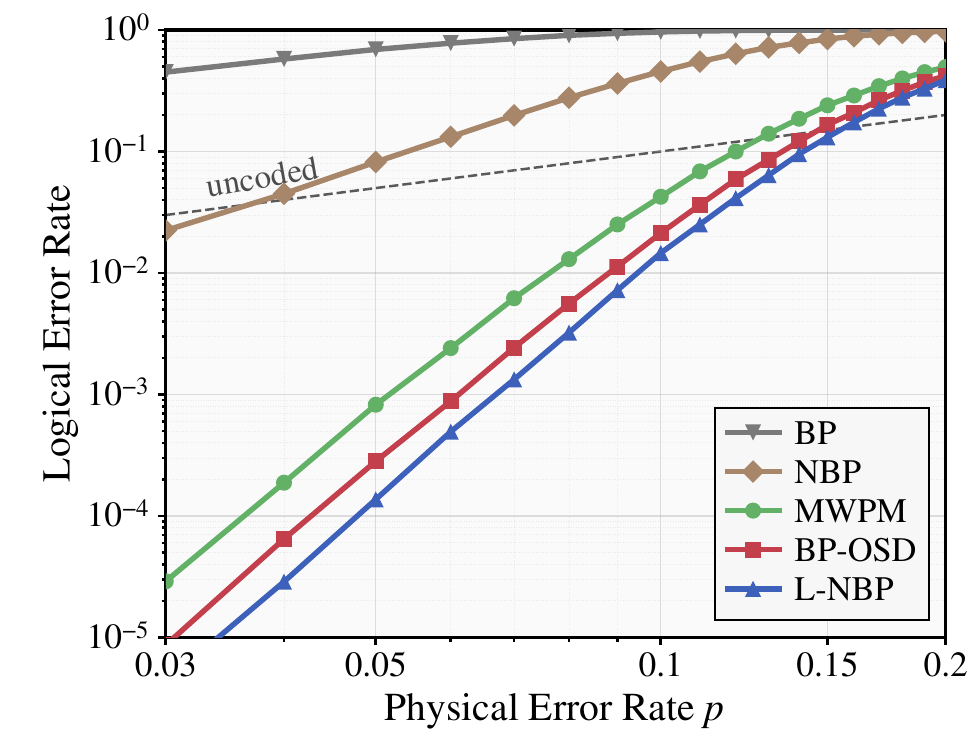}}\\
\subfloat[$d=9$, phenomenological]{\includegraphics[width=0.88\columnwidth]{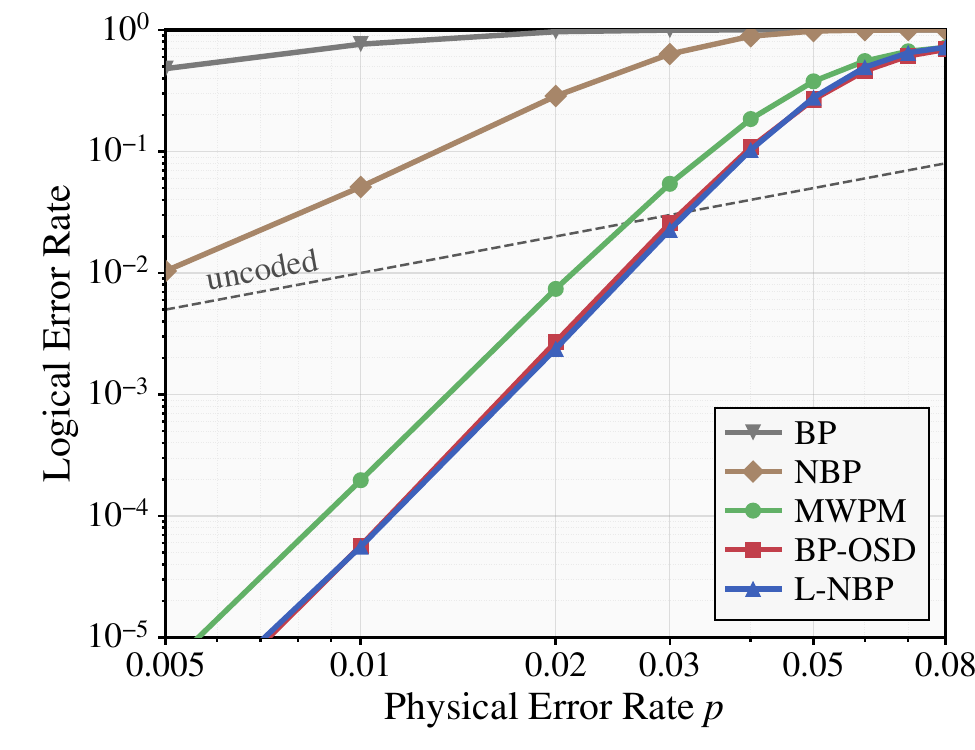}}\\
\subfloat[$d=9$, circuit-level]{\includegraphics[width=0.88\columnwidth]{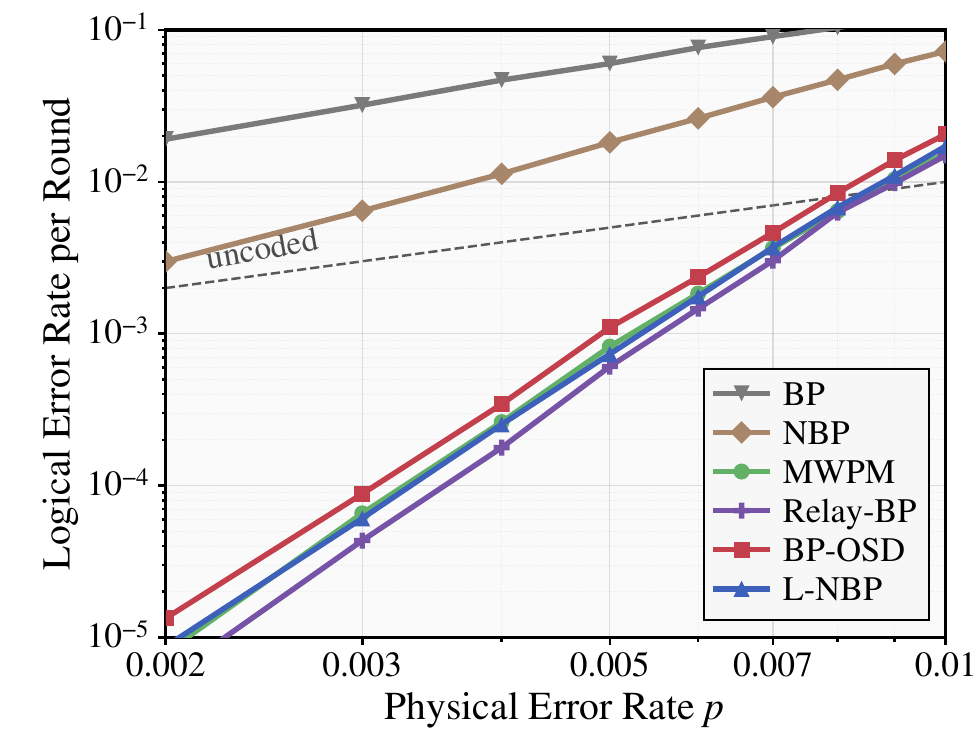}}
\caption{LER comparison under (a) code-capacity noise at $d=13$,
(b) phenomenological noise at $d=9$, and (c) circuit-level noise at $d=9$.}
\label{Fig:LER_Comparison}
\end{figure}

L-NBP, by contrast, does not need to identify which faults occurred: its NBP
module serves only to extract a soft syndrome. It can therefore operate on the
extended stabilizer matrix $\mathbf{S}_{\mathrm{ext}}$ of
\eqref{Eq:Extended_S_Matrix} rather than on $\mathbf{H}_{\mathrm{DEM}}$. The two
matrices have the same number of rows, but $\mathbf{H}_{\mathrm{DEM}}$ has far
more columns, since it introduces one column per circuit-level fault mechanism.
Running message passing on the much smaller
$\mathbf{S}_{\mathrm{ext}}$ therefore yields a considerably lighter BP stage.
Aside from obtaining the detectors $\{\mathbf{d}_r\}_{r=0}^{R}$ from Stim, L-NBP under circuit-level noise follows
exactly the same procedure as under phenomenological noise: the soft syndrome
is extracted with \eqref{Eq:MS_Round} and passed to the same logical
classifier. Further details of the circuit-level model are provided in
Appendix~\ref{app_circuit}.

\begin{figure*}[!t]
\centering
\subfloat[]{\includegraphics[width=0.34\textwidth]{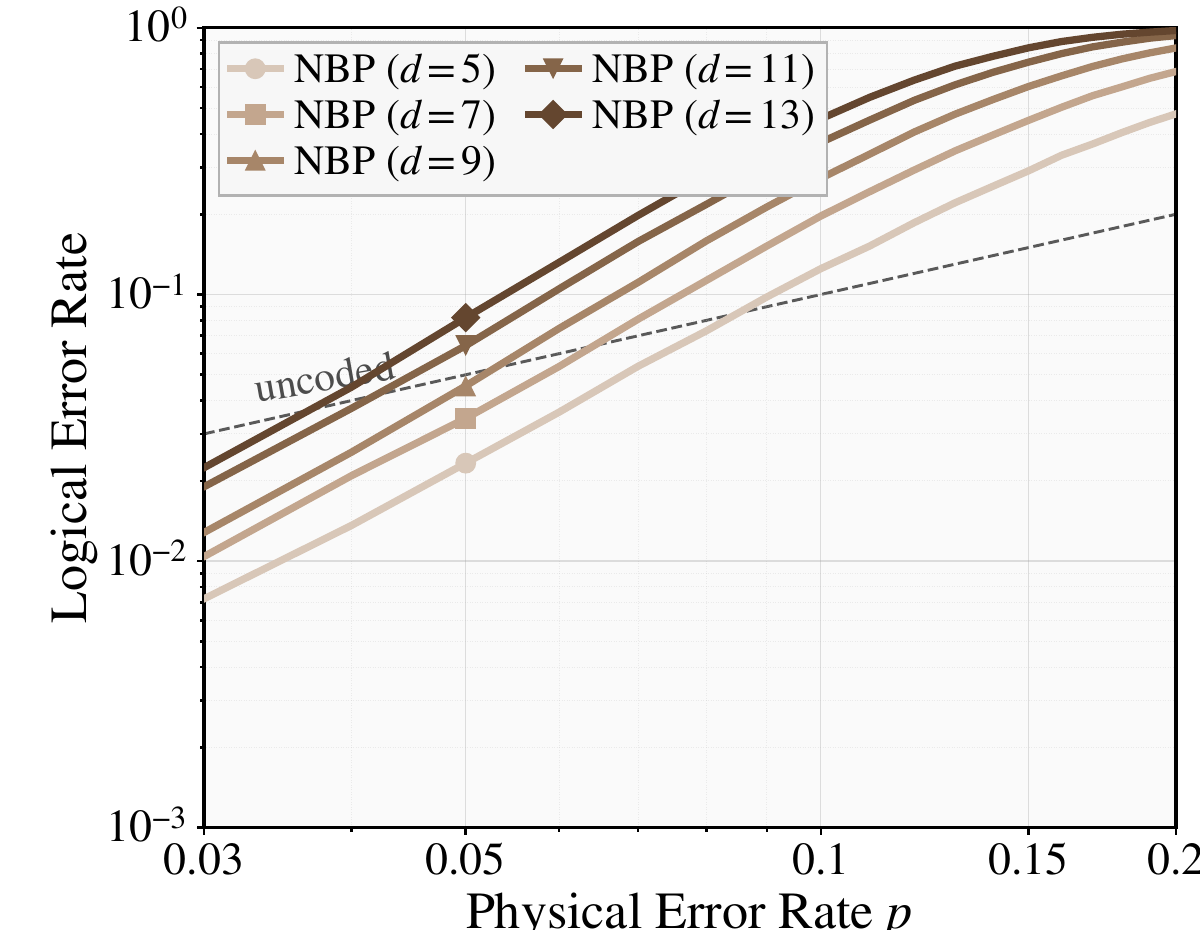}}
\subfloat[]{\includegraphics[width=0.34\textwidth]{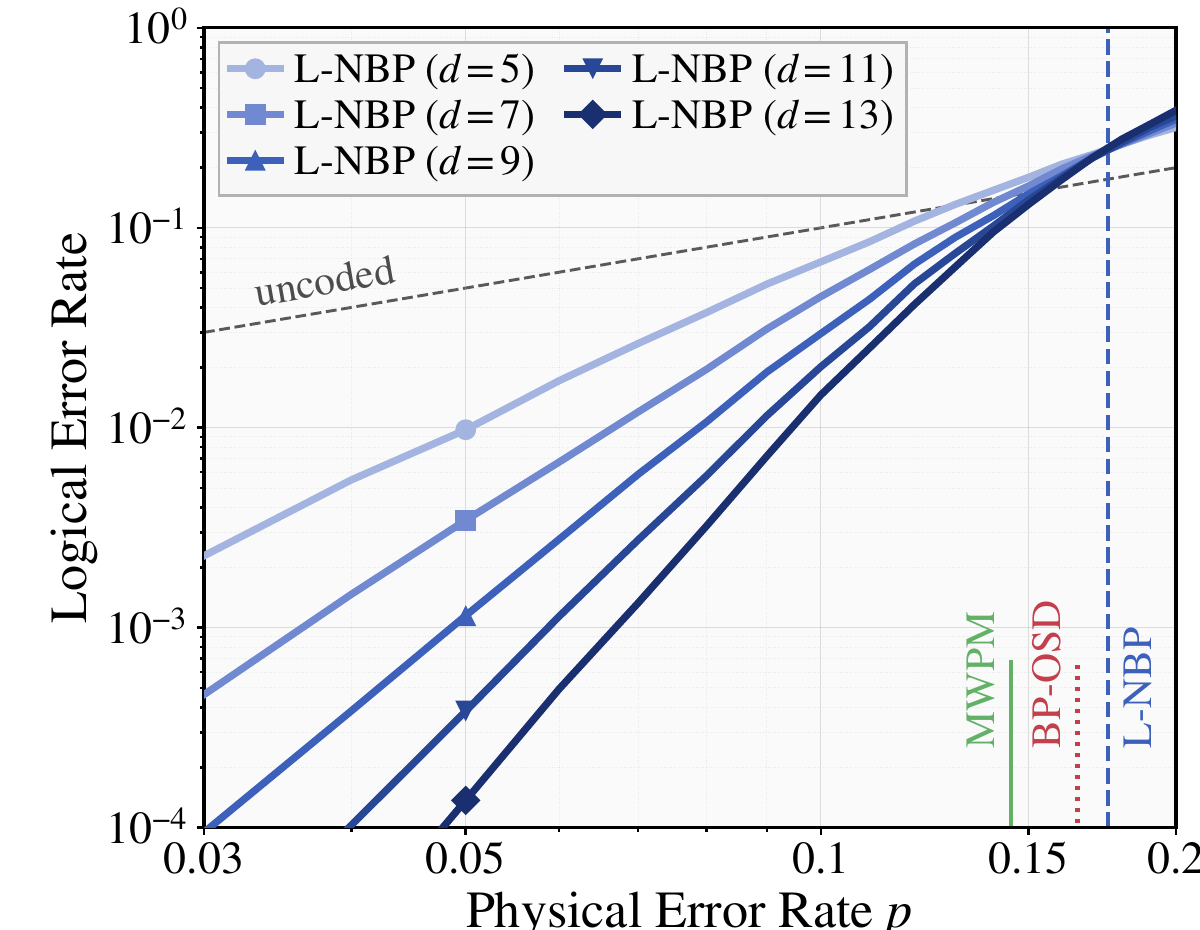}}
\subfloat[]{\includegraphics[width=0.28\textwidth]{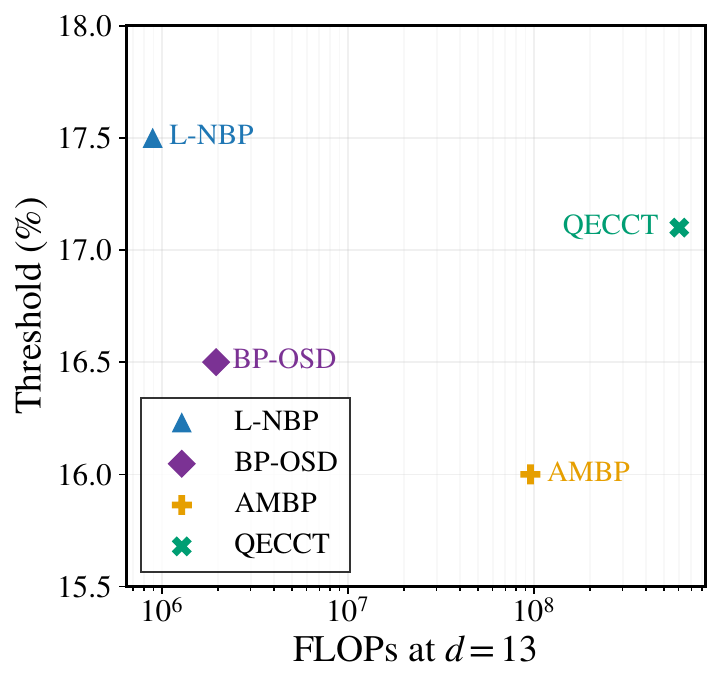}}
\caption{LER versus physical error rate across code distances for (a) NBP and
(b) L-NBP under the code-capacity model. NBP degrades as $d$ grows and shows no
threshold, whereas L-NBP exhibits clear threshold behavior with a threshold
of $17.5\%$. (c) FLOPs versus threshold at
$d=13$ for AMBP~\cite{KuoLai2022}, QECCT~\cite{Choukroun2024}, BP-OSD~\cite{PanteleevKalachev2021,RWBC2020}, and L-NBP; L-NBP attains the best
accuracy--complexity trade-off.}
\label{Fig:LNBP_Threshold}
\end{figure*}

\section{Simulation Results}
\label{sec_experiments}

\subsection{LER Comparison}
\label{subsec_ler_comparison}

Figs.~\ref{Fig:LER_Comparison}(a)--(c) compare the LER of
MWPM~\cite{Edmonds1965,Dennis2002,Fowler2012}, BP~\cite{Poulin2008},
NBP~\cite{LiuPoulin2019}, BP-OSD~\cite{PanteleevKalachev2021,RWBC2020}, and
L-NBP under the code-capacity, phenomenological, and circuit-level noise
models, respectively. MWPM is implemented with the PyMatching library \cite{Higgott2022}. BP and
NBP both use $60$ BP iterations, and NBP is trained for physical-level
decoding. BP-OSD uses OSD-0 together with a weighted min-sum BP of $60$
iterations and a fixed scaling factor of
$0.625$~\cite{PanteleevKalachev2021,RWBC2020}.

For the code-capacity model in Fig.~\ref{Fig:LER_Comparison}(a), BP performs
poorly because of short cycles and degeneracy, and NBP improves on it by
introducing trainable weights, yet still falls short of the MWPM and BP-OSD
baselines. L-NBP, in contrast, outperforms BP-OSD. The large gap between NBP
and L-NBP confirms that the gain comes primarily from changing the decoding
objective from physical to logical. The same trend holds under the
phenomenological and circuit-level noise models in
Figs.~\ref{Fig:LER_Comparison}(b) and (c): L-NBP matches or exceeds BP-OSD and
MWPM.

Under the circuit-level model, we compare in terms of LER per round and additionally compare against
Relay-BP~\cite{Muller2025}, which exploits all detector information and runs on
$\mathbf{H}_{\mathrm{DEM}}$. It employs a first stage (termed a \emph{leg}) of
$80$ BP iterations followed by $300$ additional stages of $60$ iterations each,
amounting to a maximum of $18{,}080$ iterations. In contrast, L-NBP uses only
$60$ iterations, yet it is only slightly worse than Relay-BP. As the next
subsection shows, L-NBP nonetheless holds a clear advantage in decoding
complexity.

Fig.~\ref{Fig:LNBP_Threshold} plots the LER of NBP and L-NBP across code
distances under the code-capacity model. NBP degrades as $d$ increases and shows no
threshold. L-NBP, by contrast, exhibits clear threshold behavior with a
threshold of $17.5\%$, exceeding those of MWPM ($14.5\%$) and BP-OSD
($16.5\%$). Fig.~\ref{Fig:LNBP_Threshold}(c) compares the threshold across
several decoders\footnote{MWPM is omitted from
the FLOPs comparison, as its combinatorial matching is not directly measurable in FLOPs.} together with their decoding complexity, measured in
floating-point operations (FLOPs) at $d=13$. Beyond BP-OSD, we additionally compare against the QECCT~\cite{Choukroun2024} and AMBP~\cite{KuoLai2022}. The QECCT, a
fully neural network using a transformer architecture, attains a threshold of
$17.1\%$, but its reliance on a transformer drives the cost up to $604$\,M FLOPs.
AMBP repeatedly invokes
BP---$51$ BP stages in total, each with $150$ iterations---so
that, even without a post-processing stage, it still requires $95$\,M FLOPs.
BP-OSD~\cite{PanteleevKalachev2021,RWBC2020} reaches a $16.5\%$ threshold at $1.95$\,M FLOPs, whereas L-NBP achieves a
higher $17.5\%$ threshold at only $0.89$\,M FLOPs, yielding the best
accuracy--complexity trade-off among the compared decoders.

\begin{figure*}[!t]
\centering
\subfloat[Code-capacity]{\includegraphics[width=0.33\textwidth]{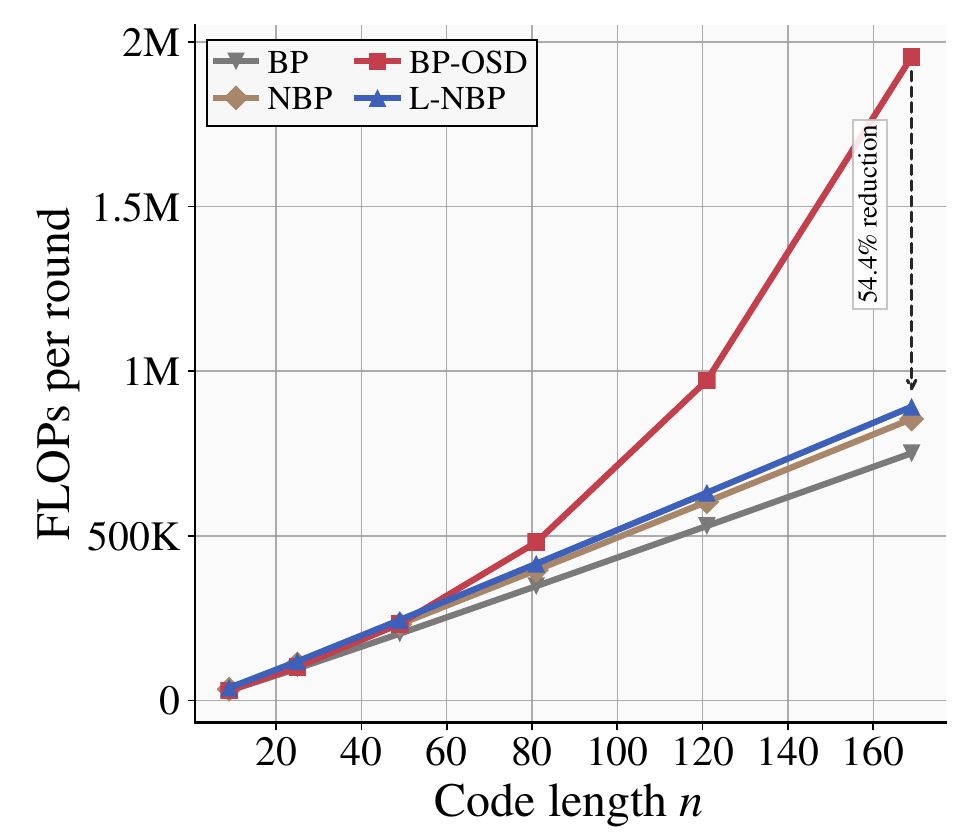}}
\subfloat[Phenomenological]{\includegraphics[width=0.33\textwidth]{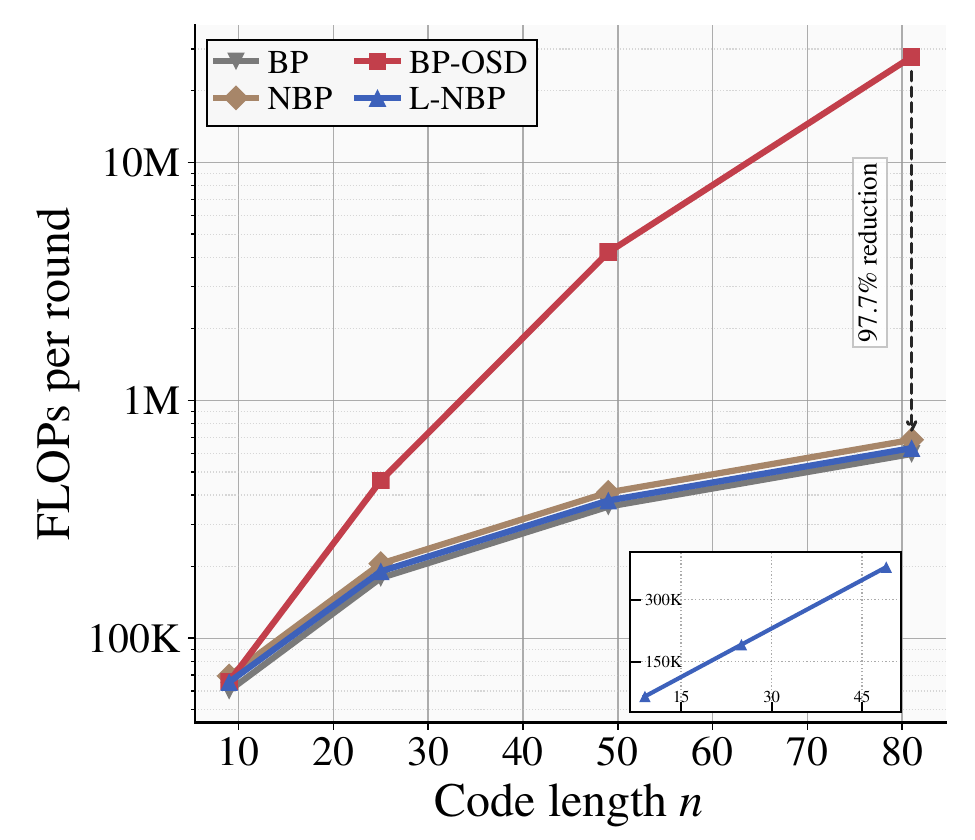}}
\subfloat[Circuit-level]{\includegraphics[width=0.33\textwidth]{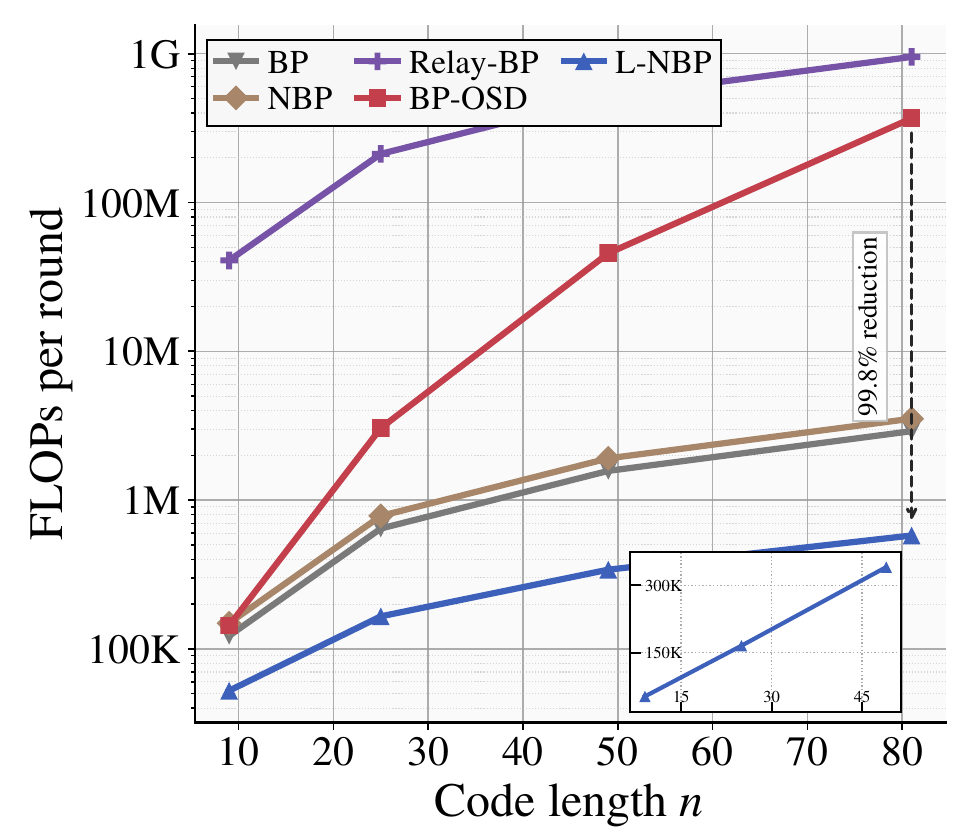}}
\caption{FLOPs versus code length for BP, NBP, BP-OSD, and L-NBP.
(a) Code-capacity: L-NBP is linear and close to NBP, with only a slight
classifier overhead. (b) Phenomenological: the gap to BP-OSD widens; the
linear-$y$ inset shows L-NBP stays linear. (c) Circuit-level: L-NBP remains
linear, while the other decoders operate on the large
$\mathbf{H}_{\mathrm{DEM}}$ and scale steeply.}
\label{Fig:Flops_by_Length}
\end{figure*}

\subsection{Complexity Analysis}

Next, we evaluate the decoding complexity in terms of the code length $n$.
Under the code-capacity model, the complexity of BP and NBP is proportional to
the number of edges in the decoding graph. Since this number grows as
$\mathcal{O}(n)$ for surface codes, both BP and NBP scale linearly in $n$. On
top of NBP, L-NBP adds the logical classifier, whose soft-syndrome extraction
and MLP are also linear in $n$, so it incurs only a slight overhead over NBP.
L-NBP thus preserves the overall linear complexity, as confirmed by
Fig.~\ref{Fig:Flops_by_Length}(a). In contrast, BP-OSD grows cubically as
$\mathcal{O}(n^3)$; already at $d=13$ ($n=169$), L-NBP achieves about a $54.4\%$
reduction in FLOPs relative to BP-OSD.

Under the phenomenological model, the decoding problem has input size
$m(d+1)=(n-1)(\sqrt{n}+1)$ for $R=d$ measurement rounds. BP and NBP therefore
incur a total cost of $\mathcal{O}(n\sqrt{n})$. In L-NBP, extracting the soft
syndrome likewise costs $\mathcal{O}(n\sqrt{n})$, and the resulting soft
syndrome of length $m = n-1$ is passed to the MLP, which costs $\mathcal{O}(n)$
over the $R$ rounds. Thus, across all rounds, BP, NBP, and L-NBP all cost
$\mathcal{O}(n\sqrt{n})$ in total, corresponding to a linear $\mathcal{O}(n)$
complexity per round. In contrast, BP-OSD costs $\mathcal{O}(n^{4.5})$ in total
and $\mathcal{O}(n^4)$ per round, as its OSD post-processing overhead grows
rapidly. As shown in Fig.~\ref{Fig:Flops_by_Length}(b), the FLOPs gap between
the two therefore widens as $n$ grows, and at $d=9$ ($n=81$), L-NBP consumes
only about $2.3\%$ of the FLOPs required by BP-OSD. The linear-$y$ inset, which
replots L-NBP alone, makes its linear $\mathcal{O}(n)$ scaling explicit.

Finally, under the circuit-level model, the BP-based decoders run over the DEM
parity-check matrix $\mathbf{H}_{\mathrm{DEM}}$, whose column size $N_f$ is very
large. As a result, the OSD stage of BP-OSD costs $\mathcal{O}(N_f^3)$, which
grows rapidly with $N_f$. L-NBP instead operates on the far more compact
extended stabilizer matrix $\mathbf{S}_{\mathrm{ext}}$, and therefore performs
the same operations as in the phenomenological case, retaining the linear
$\mathcal{O}(n)$ complexity per round. Table~\ref{tab:flops_comparison} shows
that $\mathbf{S}_{\mathrm{ext}}$ is a $720 \times 1530$ matrix for $d=9$,
compared with the much larger $720 \times 12750$ $\mathbf{H}_{\mathrm{DEM}}$.
Consequently, BP-OSD incurs a high complexity not only in its OSD
post-processing but also in its BP stage, whereas L-NBP remains lightweight in
both its BP stage and its post-processing stage (i.e., the logical classifier).
Overall, L-NBP operates at only $0.2\%$ of the complexity of BP-OSD, a $99.8\%$
reduction. Moreover, although Relay-BP has no post-processing stage, its BP
stage alone is highly complex, and L-NBP runs at only about $0.06\%$ of its
complexity. Fig.~\ref{Fig:Flops_by_Length}(c) shows that L-NBP is an order of
magnitude less complex than BP-OSD and Relay-BP, and is even lighter than plain
BP and NBP, which run on $\mathbf{H}_{\mathrm{DEM}}$. As the linear-$y$ inset
confirms, L-NBP retains its linear $\mathcal{O}(n)$ scaling under the
circuit-level noise model.

\begin{table}[t]
    \centering
    \caption{Comparison, at $d=9$, of the matrix used for BP decoding, the
    number of BP iterations, and the per-round FLOPs of the BP and
    post-processing (OSD or logical classifier) stages.}
    \label{tab:flops_comparison}
    \setlength{\tabcolsep}{4pt}
    \resizebox{\columnwidth}{!}{%
\begin{tabular}{llrrrr}
        \toprule
        \multirow{2}{*}{Method}
        & \multirow{2}{*}{Matrix}
        & \multirow{2}{*}{Iter.}
        & \multicolumn{3}{c}{FLOPs per round} \\
        \cmidrule(lr){4-6}
        & & & BP & Post & Total \\
        \midrule

        BP-OSD
        & \begin{tabular}[c]{@{}c@{}}
            $\mathbf{H}_{\mathrm{DEM}}$ \\
            $(720 \times 12750)$
          \end{tabular}
        & $60$
        & $3\,\mathrm{M}$
        & $366\,\mathrm{M}$
        & $369\,\mathrm{M}$ \\

        \addlinespace

        Relay-BP
        & \begin{tabular}[c]{@{}c@{}}
            $\mathbf{H}_{\mathrm{DEM}}$ \\
            $(720 \times 12750)$
          \end{tabular}
        & $18080$
        & $953\,\mathrm{M}$
        & $0$
        & $953\,\mathrm{M}$ \\

        \addlinespace

        L-NBP
        & \begin{tabular}[c]{@{}c@{}}
            $\mathbf{S}_{\mathrm{ext}}$ \\
            $(720 \times 1530)$
          \end{tabular}
        & $60$
        & $566\,\mathrm{K}$
        & $12\,\mathrm{K}$
        & $0.58\,\mathrm{M}$ \\

        \bottomrule
    \end{tabular}%
}
\end{table}

\section{Conclusion}
\label{sec_conclusion}
Despite its popularity as a standard decoder in classical coding, BP decoding has
not been used on its own for quantum codes because it struggles to infer the
underlying physical error. To address this, we proposed L-NBP, which combines NBP
with a logical classifier to infer the logical operator rather than the physical
error. It exploits a property unique to quantum codes: recovering
the logical operator, rather than the exact physical error, is sufficient for
correction.
Across a range of noise models, L-NBP maintains the linear complexity $\mathcal{O}(n)$ of BP decoding while matching or outperforming the
superlinear-complexity MWPM and BP-OSD decoders. Under the code-capacity model,
L-NBP attains the highest threshold of $17.5\%$ among BP-OSD, QECCT, and AMBP
while incurring the lowest complexity, achieving the best
accuracy--complexity trade-off. Compared to BP-OSD, L-NBP reduces complexity by
$99.8\%$ under the circuit-level model at $d=9$, since its BP stage only
generates a soft syndrome rather than solving the multi-round fault-estimation
task, allowing it to operate on a much smaller graph. L-NBP thus raises the
potential of BP-based decoding for QEC, achieving high accuracy, linear
complexity, and scalability all at once.

\section*{Author contributions}
Hee-Youl Kwak led the research and manuscript preparation.
Seong-Joon Park contributed to the experimental validation.
Dae-Young Yun and Eliya Nachmani provided feedback and contributed to the
revision and improvement of the manuscript. Jae-Won Kim contributed to
mathematical verification and manuscript revision.

AI tools were used to assist with simulation-code development and manuscript
review and editing. The authors remain responsible for the scientific content
and results.

\appendix

\section{BP and NBP Message Update Rules}
\label{app_bp}

For each VN $v_j$ and error type $e \in \{X, Y, Z\}$, the initial prior LLR
is set according to the noise model. For physical error rate $p$,
each Pauli error occurs on a physical qubit with probability $p/3$ under depolarizing noise, giving
\begin{equation}
    \Lambda_{j,e}^{(0)} = \ln \left( \frac{P(E_j = I)}
    {P(E_j = e)} \right) = \ln \left( \frac{1 - p}{p/3} \right).
\end{equation}
Since the exact value of $p$ is often unavailable and
minor deviations have negligible impact~\cite{Miao2025}, we use a fixed
initialization $p=0.1$.

At each iteration $\ell$ from $1$ to $\overline{\ell}$, messages are exchanged
between VNs and CNs. Let $\mu_{j \to i, e}^{(\ell)}$ denote the message from
VN $v_j$ to CN $c_i$ for error type $e$, initialized as
$\mu_{j \to i, e}^{(0)} = \Lambda_{j,e}^{(0)}$. From the previous VN message
vector
$(\mu_{j \to i, X}^{(\ell-1)}, \mu_{j \to i, Y}^{(\ell-1)}, \mu_{j \to i, Z}^{(\ell-1)})$,
we first compute a scalar VN message $\lambda_{j \to i}^{(\ell-1)}$, which
represents the belief that the error $E_j$ commutes with $S_{i,j}$:
\begin{align}
\label{Eq:Belief_Quantization}
    \lambda_{j \to i}^{(\ell-1)}
    &= \ln \left(
        \frac{1 + \exp\!\left(-\mu_{j \to i,\, S_{i,j}}^{(\ell-1)}\right)}
             {\exp\!\left(-\mu_{j \to i,\, e_1}^{(\ell-1)}\right)
              + \exp\!\left(-\mu_{j \to i,\, e_2}^{(\ell-1)}\right)}
       \right) \\\nonumber
    &= \zeta\!\left(-\mu_{j \to i,\, S_{i,j}}^{(\ell-1)}\right)
    + \mu_{j \to i,\, e_1}^{(\ell-1)}
    \\\nonumber
    &\quad{}- \zeta\!\left(\mu_{j \to i,\, e_1}^{(\ell-1)}
    - \mu_{j \to i,\, e_2}^{(\ell-1)}\right),
\end{align}
where $\{e_1,e_2\}=\{X,Y,Z\}\setminus\{S_{i,j}\}$ and
$\zeta(x) = \ln(1 + e^x)$ is the softplus function.

On the CN side, the CN message $\nu_{i \to j}^{(\ell)}$ from CN $c_i$ to VN
$v_j$ is computed via the min-sum rule, incorporating the measured syndrome
bit $s_i$:
\begin{equation}
    \label{Eq:BP_CtV}
    \begin{aligned}
    \nu_{i \to j}^{(\ell)} &= (-1)^{s_i}
    \!\left( \prod_{j' \in \mathcal{N}_c(i) \setminus \{j\}}
    \!\!\!\!\!\!\mathrm{sgn}\!\left(\lambda_{j' \to i}^{(\ell-1)}\right) \right)\!
    \\
    &\quad{}\times \min_{j' \in \mathcal{N}_c(i) \setminus \{j\}}\!
    \left|\lambda_{j' \to i}^{(\ell-1)}\right|.
    \end{aligned}
\end{equation}
The VN message for each error type $e$ is then updated as a weighted,
extrinsic combination of the prior LLR and the incoming CN messages,
followed by a damping step:
\begin{align}
    \hat{\mu}_{j \to i,\, e}^{(\ell)} &= \beta_{j}^{(\ell)}\,
    \Lambda_{j,e}^{(0)}
    + \!\!\sum_{\substack{i' \in \mathcal{N}_v(j) \setminus \{i\} \\
    \langle e,\, S_{i',j} \rangle = 1}}\!\!
    \alpha_{i' \to j}^{(\ell)}\, \nu_{i' \to j}^{(\ell)},
    \label{Eq:Weighted_Sum}\\
    \mu_{j \to i,\, e}^{(\ell)} &= \hat{\mu}_{j \to i,\, e}^{(\ell)}
    + (1-\eta_{j \to i}^{(\ell)})\, \mu_{j \to i,\, e}^{(\ell-1)}.
    \label{Eq:Damping}
\end{align}
Here, $\alpha_{i' \to j}^{(\ell)}$, $\beta_{j}^{(\ell)}$, and
$\eta_{j \to i}^{(\ell)}$ are trainable weights:
\begin{equation}
    \theta_{\mathrm{N}} = \bigl\{\alpha_{i \to j}^{(\ell)},\,\beta_{j}^{(\ell)},\,
    \eta_{j \to i}^{(\ell)}\bigr\}.
    \label{Eq:NBP_Params}
\end{equation} In the extrinsic sum of \eqref{Eq:Weighted_Sum}, only CNs whose stabilizers
anticommute with the error type $e$ (i.e., $\langle e, S_{i',j}\rangle = 1$)
contribute. The damping step
\eqref{Eq:Damping} mixes the updated message with the previous-iteration
message to improve convergence on graphs with short cycles.

At each iteration, the posterior LLR $\Lambda_{j,e}^{(\ell)}$ is computed
analogously to \eqref{Eq:Weighted_Sum}, but over the full neighborhood
$\mathcal{N}_v(j)$:
\begin{equation}
    \Lambda_{j,e}^{(\ell)} = \beta_{j}^{(\ell)}\,
    \Lambda_{j,e}^{(0)}
    + \sum_{\substack{i' \in \mathcal{N}_v(j)\\
    \langle e,\, S_{i',j} \rangle = 1}}
    \alpha_{i' \to j}^{(\ell)}\, \nu_{i' \to j}^{(\ell)}.
    \label{Eq:Post_Sum}
\end{equation}

In classical BP, all of the weights in $\theta_{\rm N}$ are fixed to one. Weighted BP
instead applies a single fixed scaling factor to $\alpha_{i \to j}^{(\ell)}$.

\begin{figure*}[!t]
\centering
\subfloat[]{\includegraphics[width=0.24\textwidth]{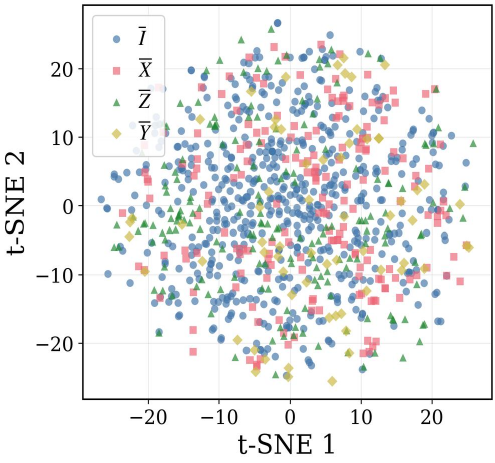}}
\subfloat[]{\includegraphics[width=0.24\textwidth]{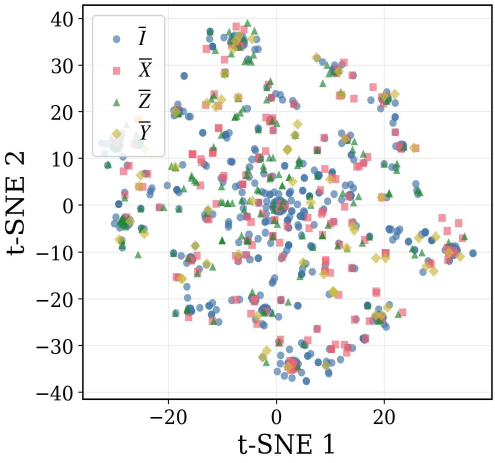}}
\subfloat[]{\includegraphics[width=0.24\textwidth]{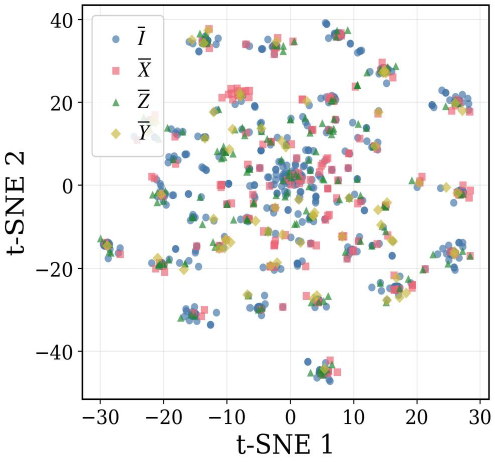}}
\subfloat[]{\includegraphics[width=0.24\textwidth]{tsne_LNBP.pdf}}
    \caption{t-SNE visualization of (a) the measured hard syndrome
    $\mathbf{s}$, (b) the soft syndrome from untrained BP, (c) the soft
    syndrome from independently trained NBP, and (d) the E2E-trained L-NBP soft
    syndrome $\tilde{\mathbf{s}}$ (surface code $d=9$, $p=0.05$). Only the E2E-trained L-NBP soft
    syndrome forms compact, well-separated clusters by logical operator.}
\label{Fig:t_SNE_ablation}
\end{figure*}

\section{Training and Evaluation Details}
\label{app_training}

For training, we use the Adam optimizer~\cite{KingmaBa2015} with a batch
size of $256$ and $1{,}000$ batches per epoch, for a total of $1000$ epochs.
The learning rate is set to $10^{-4}$ and annealed by a factor of $1/100$
with a cosine schedule. All trainable weights are initialized using Xavier
initialization~\cite{Glorot2010}. Training samples are collected at
$p=0.15$, $0.03$, and $0.007$ for the code-capacity, phenomenological, and
circuit-level models, respectively.
The decoder parameters are set to $\overline{\ell}=60$
BP iterations, a hidden dimension of $h=256$ for the logical
classifier, and an iteration sampling interval of $\Delta=10$.
Training is performed on a single NVIDIA
RTX A5000 GPU; for the $d=9$ surface code under the code-capacity model, one
epoch of L-NBP training takes about $30$ seconds, and the full training run
takes roughly $8$ hours.

Every LER estimate reported in Figs.~\ref{Fig:LER_Comparison}
and~\ref{Fig:LNBP_Threshold} is obtained from at least $10^5$ Monte Carlo
trials, with the simulation continued until at least $1000$ logical failures are
collected, which yields statistically reliable estimates. The threshold is
defined as the crossing point among the LER curves for different distances.

\section{Circuit-Level Noise Model Details}
\label{app_circuit}

We use Stim's standard rotated surface-code memory circuit, \texttt{surface\_code:rotated\_memory\_z}, with code distance $d$ and $R=d$ syndrome-extraction rounds, followed by a final data-qubit measurement. A depolarizing error of probability $p$ is applied after each Clifford gate and to the data qubits before each syndrome round. Thus, each non-identity single- and two-qubit Pauli error occurs with probability $p/3$ and $p/15$, respectively. The reset and measurement flip probabilities are set to $p$. 

For the $Z$-memory experiment, the $X$-type stabilizer outcomes at the first
and final temporal boundaries are not deterministic and therefore do not
define detectors in Stim. Consequently, Stim reports $Rm$ detectors rather than $(R+1)m$. Accordingly, when constructing $\mathbf{S}_{\mathrm{ext}}$ for L-NBP, we trim the corresponding boundary rows so that its detector rows match those reported by Stim. Thus, both $\mathbf{H}_{\mathrm{DEM}}$ and the trimmed $\mathbf{S}_{\mathrm{ext}}$ operate on the same set of $Rm$ detectors. A logical failure is declared when the decoder's predicted logical-observable flip differs from that reported by Stim.

\section{Ablation Study}
\label{app_ablation}

\begin{table}[h]
\centering
\caption{Ablation results for the surface code with $d=11$ under depolarizing
noise.}
\label{tab:decoder_ablation}
\small
\setlength{\tabcolsep}{2.3pt}
\renewcommand{\arraystretch}{1.12}
\begin{tabular}{@{}l >{\raggedright\arraybackslash}p{0.46\columnwidth} c c@{}}
\toprule
\multirow{2}{*}{Method}
& \multirow{2}{*}{Input to classifier}
& E2E & LER \\
& & Train & $(p=0.05)$ \\
\midrule
Baseline
& Soft syndrome~\eqref{Eq:soft_syndrome}
& Yes & $3.8{\times}10^{-4}$ \\

Variant A
& Measured hard syndrome $\mathbf{s}$
& No & $2.3{\times}10^{-2}$ \\

Variant B
& Residual hard syndrome $\hat{\mathbf{s}}\oplus\mathbf{s}$
& No & $4.5{\times}10^{-1}$ \\

Variant C
& Soft syndrome~\eqref{Eq:soft_syndrome}
& No & $3.5{\times}10^{-1}$ \\

Variant D
& Soft syndrome~\eqref{Eq:Message_based_soft_syndrome}
& Yes & $6.1{\times}10^{-4}$ \\

Variant E
& Posterior LLRs $\{\Lambda_{j,e}^{(\overline{\ell})}\}$
& Yes & $4.0{\times}10^{-4}$ \\
\bottomrule
\end{tabular}
\end{table}

Table~\ref{tab:decoder_ablation} reports ablation results for the surface code with $d=11$. The L-NBP baseline uses the proposed soft syndrome in \eqref{Eq:soft_syndrome} with E2E training. Variant A directly feeds the measured hard syndrome $\mathbf{s}$ to the MLP, without using the NBP module. This variant therefore reduces to a standalone MLP-based logical decoder and achieves an LER of $2.3\times10^{-2}$, substantially worse than the L-NBP baseline. This result shows that directly classifying the measured syndrome with a lightweight MLP is insufficient for accurate logical decoding. 

Variant B instead feeds the residual hard syndrome $\hat{\mathbf{s}}\oplus\mathbf{s}$ to the MLP, where $\hat{\mathbf{s}}=\mathbf{S}\odot\hat{E}$ is obtained from the hard decision of the NBP module. This setting is in line with the residual-syndrome post-processing of~\cite{Meinerz2022,Chamberland2026,Zhang2025}. Because the hard decision prevents gradients from propagating through the NBP module, only the logical classifier is trained. The resulting LER remains poor, indicating that retaining only hard residual information discards information that is important for logical classification. 

Variant C uses the proposed soft syndrome but first trains the NBP module independently as a physical-level decoder and then freezes it before training the logical classifier. Its poor performance shows that simply providing soft NBP beliefs is not sufficient; the NBP module must be optimized jointly with the logical classifier so that its beliefs become informative for the logical-level objective. 

Variant D considers an alternative soft syndrome constructed directly from the edge messages $\lambda_{j\to i}^{(\ell)}$ in \eqref{Eq:Belief_Quantization}, which has previously been explored for quantum BP decoding~\cite{Raveendran2023}. In contrast to the extrinsic aggregation in \eqref{Eq:BP_CtV}, this message-based soft syndrome is computed over all incoming edge messages at each CN: \begin{equation} \left( \prod_{j \in \mathcal{N}_c(i)} \mathrm{sgn}(\lambda_{j \to i}^{(\ell)}) \right) \min_{j \in \mathcal{N}_c(i)} |\lambda_{j \to i}^{(\ell)}| \in \mathbb{R}. \label{Eq:Message_based_soft_syndrome} \end{equation} Its sign, however, is not necessarily consistent with the syndrome estimate $\hat{\mathbf{s}}$ determined by the posterior LLRs. Although this variant is trained E2E and performs substantially better than the hard-syndrome variants, its LER is still higher than that of the proposed posterior-based soft syndrome. 

Finally, Variant E directly feeds the posterior LLRs $\{\Lambda_{j,e}^{(\overline{\ell})}\}_{j,e}$ to the MLP without constructing the soft syndrome. It achieves an LER of $4.0\times10^{-4}$, comparable to the proposed L-NBP baseline. However, this approach increases the classifier input dimension from $m$ to $3n$ even in the single-round setting. The difference becomes more pronounced under multi-round measurement models, where the posterior LLRs are produced for every qubit and measurement round, causing the classifier input dimension to grow with the number of rounds.

Fig.~\ref{Fig:t_SNE_ablation} presents a t-SNE-based ablation study. The measured hard syndrome (Fig.~\ref{Fig:t_SNE_ablation}(a)) and the soft syndromes from untrained BP (Fig.~\ref{Fig:t_SNE_ablation}(b)) and independently trained NBP (Fig.~\ref{Fig:t_SNE_ablation}(c)) all remain heavily overlapped across logical operators. Only the E2E-trained soft syndrome (Fig.~\ref{Fig:t_SNE_ablation}(d)) collapses into compact, well-separated clusters.

\bibliographystyle{quantum}
\bibliography{Reference}

@PREAMBLE{
 "\providecommand{\noopsort}[1]{}" 
 # "\providecommand{\singleletter}[1]{#1}%" 
}

@ARTICLE{Shor1999,
   author       = "P. W. Shor",
   title        = "Polynomial-time algorithms for prime factorization and discrete logarithms on a quantum computer",
   journal      = "SIAM Rev.",
   volume       = "41",
   number       = "2",
   pages        = "303--332",
   year         = "1999",
   month        = "Jun",
   doi          = "10.1137/S0036144598347011"
}

@ARTICLE{Delfosse2023,
   author       = "N. Delfosse and A. Paz and A. Vaschillo and K. M. Svore",
   title        = "How to choose a decoder for a fault-tolerant quantum computer? The speed vs accuracy trade-off",
   journal      = "arXiv preprint arXiv:2310.15313",
   year         = "2023",
   month        = "Oct",
   doi          = "10.48550/arXiv.2310.15313"
}

@ARTICLE{AspuruGuzik2005,
   author       = "A. Aspuru-Guzik and A. D. Dutoi and P. J. Love and M. Head-Gordon",
   title        = "Simulated quantum computation of molecular energies",
   journal      = "Science",
   volume       = "309",
   number       = "5741",
   pages        = "1704--1707",
   year         = "2005",
   month        = "Sep",
   doi          = "10.1126/science.1113479"
}

@ARTICLE{Shor1995,
   author       = "P. W. Shor",
   title        = "Scheme for reducing decoherence in quantum computer memory",
   journal      = "Phys. Rev. A",
   volume       = "52",
   number       = "4",
   pages        = "R2493(R)--R2496(R)",
   year         = "1995",
   month        = "Oct",
   doi          = "10.1103/PhysRevA.52.R2493"
}

@ARTICLE{Steane1996,
   author       = "A. M. Steane",
   title        = "Error correcting codes in quantum theory",
   journal      = "Phys. Rev. Lett.",
   volume       = "77",
   pages        = "793--797",
   year         = "1996",
   doi          = "10.1103/PhysRevLett.77.793"
}

@ARTICLE{Gottesman1996,
  author  = "D. Gottesman",
  title   = "Class of quantum error-correcting codes saturating the quantum Hamming bound",
  journal = "Phys. Rev. A",
  volume  = "54",
  pages   = "1862--1868",
  year    = "1996",
  doi     = "10.1103/PhysRevA.54.1862"
}

@ARTICLE{GoogleQuantumAI2023,
   author       = "{Google Quantum AI}",
   title        = "Suppressing quantum errors by scaling a surface code logical qubit",
   journal      = "Nature",
   volume       = "614",
   pages        = "676--681",
   year         = "2023",
   month        = "Feb",
   doi          = "10.1038/s41586-022-05434-1"
}

@ARTICLE{GoogleQuantumAI2024,
   author       = "{Google Quantum AI}",
   title        = "Quantum error correction below the surface code threshold",
   journal      = "Nature",
   volume       = "638",  
   pages        = "920--926",  
   year         = "2025",
   doi          = "10.1038/s41586-024-08449-y"
}

@ARTICLE{Terhal2015,
   author       = "B. M. Terhal",
   title        = "Quantum error correction for quantum memories",
   journal      = "Rev. Mod. Phys.",
   volume       = "87",
   number       = "2",
   pages        = "307--346",
   year         = "2015",
   month        = "Apr",
   doi          = "10.1103/RevModPhys.87.307"
}

@ARTICLE{Gallager1962,
   author       = "R. G. Gallager",
   title        = "Low-density parity-check codes",
   journal      = "IRE Trans. Inf. Theory",
   volume       = "8",
   number       = "1",
   pages        = "21--28",
   year         = "1962",
   month        = "Jan",
   doi          = "10.1109/TIT.1962.1057683"
}

@ARTICLE{Kitaev2003,
   author       = "A. Yu. Kitaev",
   title        = "Fault-tolerant quantum computation by anyons",
   journal      = "Annals of Physics",
   volume       = "303",
   number       = "1",
   pages        = "2--30",
   year         = "2003",
   month        = "Jan",
   doi          = "10.1016/S0003-4916(02)00018-0"
}

@ARTICLE{Dennis2002,
   author       = "E. Dennis and A. Kitaev and A. Landahl and J. Preskill",
   title        = "Topological quantum memory",
   journal      = "J. Math. Phys.",
   volume       = "43",
   number       = "9",
   pages        = "4452--4505",
   year         = "2002",
   month        = "Sep",
   doi          = "10.1063/1.1499754"
}

@ARTICLE{Fowler2012,
   author       = "A. G. Fowler and M. Mariantoni and J. M. Martinis and A. N. Cleland",
   title        = "Surface codes: Towards practical large-scale quantum computation",
   journal      = "Phys. Rev. A",
   volume       = "86",
   pages        = "032324",
   year         = "2012",
   month        = "Sep",
   doi          = "10.1103/PhysRevA.86.032324"
}

@ARTICLE{MacKay2004,
   author       = "D. J. C. MacKay and G. Mitchison and P. L. McFadden",
   title        = "Sparse-graph codes for quantum error correction",
   journal      = "IEEE Trans. Inf. Theory",
   volume       = "50",
   number       = "10",
   pages        = "2315--2330",
   year         = "2004",
   doi          = "10.1109/TIT.2004.834737"
}

@ARTICLE{PanteleevKalachev2021,
   author       = "Pavel Panteleev and Gleb Kalachev",
   title        = "Degenerate Quantum LDPC Codes With Good Finite Length Performance",
   journal      = "Quantum",
   volume       = "5",
   pages        = "585",
   year         = "2021",
   month        = "Nov",
   doi          = "10.22331/q-2021-11-22-585"
}

@ARTICLE{Poulin2008,
   author       = "D. Poulin and Y. Chung",
   title        = "On the iterative decoding of sparse quantum codes",
   journal      = "Quantum Inf. Comput.",
   volume       = "8",
   pages        = "987--1000",
   year         = "2008",
   doi          = "10.26421/QIC8.10-8"
}

@ARTICLE{Babar2015,
   author       = "Z. Babar and P. Botsinis and D. Alanis and S. X. Ng and L. Hanzo",
   title        = "Fifteen years of quantum LDPC coding and improved decoding strategies",
   journal      = "IEEE Access",
   volume       = "3",
   pages        = "2492--2519",
   year         = "2015",
   doi          = "10.1109/ACCESS.2015.2503267"
}

@ARTICLE{RWBC2020,
   author       = "Joschka Roffe and David R. White and Simon Burton and Earl T. Campbell",
   title        = "Decoding Across the Quantum LDPC Code Landscape",
   journal      = "Phys. Rev. Research",
   volume       = "2",
   number       = "4",
   pages        = "043423",
   year         = "2020",
   month        = "Dec",
   doi          = "10.1103/PhysRevResearch.2.043423"
}

@ARTICLE{Varsamopoulos2017,
   author       = "S. Varsamopoulos and B. Criger and K. Bertels",
   title        = "Decoding small surface codes with feedforward neural networks",
   journal      = "Quantum Sci. Technol.",
   volume       = "3",
   number       = "1",
   pages        = "015004",
   year         = "2017",
   doi          = "10.1088/2058-9565/aa955a"
}

@ARTICLE{ChamberlandRonagh2018,
   author       = "C. Chamberland and P. Ronagh",
   title        = "Deep neural decoders for near term fault-tolerant experiments",
   journal      = "Quantum Sci. Technol.",
   volume       = "3",
   number       = "4",
   pages        = "044002",
   year         = "2018",
   doi          = "10.1088/2058-9565/aad1f7"
}

@ARTICLE{JungAliHa2024,
   author       = "Hyunwoo Jung and Inayat Ali and Jeongseok Ha",
   title        = "Convolutional Neural Decoder for Surface Codes",
   journal      = "IEEE Transactions on Quantum Engineering",
   volume       = "5",
   pages        = "3102513",
   year         = "2024",
   doi          = "10.1109/TQE.2024.3102513"
}

@ARTICLE{Bausch2024,
   author       = "Johannes Bausch and Andrew W. Senior and Francisco J. H. Heras and Thomas Edlich and Alex Davies and Michael Newman and Cody Jones and Kevin Satzinger and Murphy Yuezhen Niu and Sam Blackwell and George Holland and Dvir Kafri and Juan Atalaya and Craig Gidney and Demis Hassabis and Sergio Boixo and Hartmut Neven and Pushmeet Kohli",
   title        = "Learning high-accuracy error decoding for quantum processors",
   journal      = "Nature",
   volume       = "635",
   pages        = "834--840",
   year         = "2024",
   month        = "Nov",
   doi          = "10.1038/s41586-024-08148-8"
}

@ARTICLE{LiuPoulin2019,
   author       = "Ye-Hua Liu and David Poulin",
   title        = "Neural Belief-Propagation Decoders for Quantum Error-Correcting Codes",
   journal      = "Phys. Rev. Lett.",
   volume       = "122",
   number       = "20",
   pages        = "200501",
   year         = "2019",
   month        = "May",
   doi          = "10.1103/PhysRevLett.122.200501"
}

@ARTICLE{KuoLai2022,
   author       = "Kao-Yueh Kuo and Ching-Yi Lai",
   title        = "Exploiting degeneracy in belief propagation decoding of quantum codes",
   journal      = "npj Quantum Information",
   volume       = "8",
   number       = "1",
   pages        = "111",
   year         = "2022",
   month        = "Sep",
   doi          = "10.1038/s41534-022-00623-2"
}

@ARTICLE{Miao2025,
   author       = "Sisi Miao and Alexander Schnerring and Haizheng Li and Laurent Schmalen",
   title        = "Quaternary Neural Belief Propagation Decoding of Quantum LDPC Codes With Overcomplete Check Matrices",
   journal      = "IEEE Access",
   volume       = "11",
   pages        = "1--1",
   year         = "2025",
   month        = "Feb",
   doi          = "10.1109/ACCESS.2025.3539475"
}

@ARTICLE{Muller2025,
  author  = {M\"{u}ller, Tristan and Alexander, Thomas and Beverland, Michael E. and B\"{u}hler, Markus and Johnson, Blake R. and Maurer, Thilo and Vandeth, Drew},
  title   = {Improved belief propagation is sufficient for real-time decoding of quantum memory},
  journal = {arXiv preprint arXiv:2506.01779},
  year    = {2025},
   url = "https://arxiv.org/abs/2506.01779"
}

@ARTICLE{Raveendran2023,
   author       = "N. Raveendran and J. Valls and A. K. Pradhan and N. Rengaswamy and F. Garcia-Herrero and B. Vasi{\'c}",
   title        = "Soft syndrome iterative decoding of quantum {LDPC} codes and hardware architectures",
   journal      = "EPJ Quantum Technol.",
   volume       = "10",
   number       = "1",
   pages        = "1--24",
   year         = "2023",
   month        = "Oct",
   doi          = "10.1140/epjqt/s40507-023-00201-1"
}

@INPROCEEDINGS{KuoChernLai2021,
   author       = "K.-Y. Kuo and I.-C. Chern and C.-Y. Lai",
   title        = "Decoding of quantum data-syndrome codes via belief propagation",
   booktitle    = "Proc. IEEE Int. Symp. Inf. Theory (ISIT)",
   pages        = "1552--1557",
   year         = "2021",
   month        = "Jul",
   doi          = "10.1109/ISIT45174.2021.9518018"
}

@ARTICLE{KuoLai2025,
   author       = "K.-Y. Kuo and C.-Y. Lai",
   title        = "Generalized quantum data-syndrome codes and belief propagation decoding for phenomenological noise",
   journal      = "IEEE Trans. Inf. Theory",
   volume       = "71",
   number       = "3",
   pages        = "1824--1840",
   year         = "2025",
   month        = "Mar",
   doi          = "10.1109/TIT.2025.3528878"
}

@ARTICLE{Edmonds1965,
   author       = "J. Edmonds",
   title        = "Paths, trees, and flowers",
   journal      = "Can. J. Math.",
   volume       = "17",
   pages        = "449--467",
   year         = "1965",
   doi          = "10.4153/CJM-1965-045-4"
}

@INPROCEEDINGS{Choukroun2024,
   author       = "Y. Choukroun and L. Wolf",
   title        = "Deep quantum error correction",
   booktitle    = "Proc. AAAI Conf. Artif. Intell. (AAAI)",
   volume       = "38",
   pages        = "64--72",
   year         = "2024",
   month        = "Mar",
   doi          = "10.1609/aaai.v38i1.27756"
}

@ARTICLE{Park2026,
   author       = "S.-J. Park and H.-Y. Kwak and Y. Kim",
   title        = "Hierarchical Qubit-Merging Transformer for Quantum Error Correction",
   journal      = "arXiv preprint",
   eprint       = "2510.11593",
   archivePrefix= "arXiv",
   primaryClass = "quant-ph",
   year         = "2025",
   month        = "Mar",
   url          = "https://arxiv.org/abs/2510.11593"
}

@ARTICLE{VanDerMaaten2008,
   author       = "L. van der Maaten and G. Hinton",
   title        = "Visualizing data using t-{SNE}",
   journal      = "J. Mach. Learn. Res.",
   volume       = "9",
   number       = "11",
   pages        = "2579--2605",
   year         = "2008",
   month        = "Nov",
   doi          = "10.1145/1390681.1390693"
}

@ARTICLE{Higgott2022,
   author       = "O. Higgott",
   title        = "{PyMatching}: A Python Package for Decoding Quantum Codes with Minimum-Weight Perfect Matching",
   journal      = "ACM Trans. Quantum Comput.",
   volume       = "3",
   number       = "3",
   pages        = "16",
   year         = "2022",
   month        = "Jun",
   doi          = "10.1145/3505637"
}

@ARTICLE{Higgott2023,
   author       = "O. Higgott and T. C. Bohdanowicz and A. Kubica and S. T. Flammia and E. T. Campbell",
   title        = "Improved decoding of circuit noise and fragile boundaries of tailored surface codes",
   journal      = "Phys. Rev. X",
   volume       = "13",
   number       = "3",
   pages        = "031007",
   year         = "2023",
   month        = "Jul",
   doi          = "10.1103/PhysRevX.13.031007"
}

@ARTICLE{Kwak2025,
   author       = "H.-Y. Kwak and S.-J. Park and H. Jung and J. Ha and J.-W. Kim",
   title        = "Evolutionary {BP+OSD} Decoding for Low-Latency Quantum Error Correction",
   journal      = "arXiv preprint",
   eprint       = "2512.18273",
   archivePrefix= "arXiv",
   primaryClass = "quant-ph",
   year         = "2025",
   month        = "Dec",
   url          = "https://arxiv.org/abs/2512.18273"
}

@ARTICLE{Chamberland2026,
   author       = "C. Chamberland and J. Olle and M. Li and S. Thornton and I. Baratta",
   title        = "Fast and accurate {AI}-based pre-decoders for surface codes",
   journal      = "arXiv preprint",
   eprint       = "2604.12841",
   archivePrefix= "arXiv",
   primaryClass = "quant-ph",
   year         = "2026",
   url          = "https://arxiv.org/abs/2604.12841"
}

@ARTICLE{Zhang2025,
   author       = "Kai Zhang and Jubo Xu and Fang Zhang and Linghang Kong and Zhengfeng Ji and Jianxin Chen",
   title        = "{LATTE}: A decoding architecture for quantum computing with temporal and spatial scalability",
   journal      = "arXiv preprint",
   eprint       = "2509.03954",
   archivePrefix= "arXiv",
   primaryClass = "quant-ph",
   year         = "2025",
   month        = "Sep",
   url          = "https://arxiv.org/abs/2509.03954"
}

@ARTICLE{Meinerz2022,
   author       = "K. Meinerz and C.-Y. Park and S. Trebst",
   title        = "Scalable Neural Decoder for Topological Surface Codes",
   journal      = "Phys. Rev. Lett.",
   volume       = "128",
   number       = "8",
   pages        = "080505",
   year         = "2022",
   month        = "Feb",
   doi          = "10.1103/PhysRevLett.128.080505"
}

@INPROCEEDINGS{Glorot2010,
   author       = "X. Glorot and Y. Bengio",
   title        = "Understanding the difficulty of training deep feedforward neural networks",
   booktitle    = "Proc. Int. Conf. Artif. Intell. Stat. (AISTATS)",
   pages        = "249--256",
   year         = "2010",
   url = "https://proceedings.mlr.press/v9/glorot10a.html"
}

@INPROCEEDINGS{Grover1996,
   author       = "L. K. Grover",
   title        = "A fast quantum mechanical algorithm for database search",
   booktitle    = "Proc. 28th Annu. ACM Symp. Theory Comput. (STOC)",
   pages        = "212--219",
   year         = "1996",
   month        = "Jul",
   doi          = "10.1145/237814.237866"
}

@ARTICLE{HarrowHassidimLloyd2009,
   author       = "A. W. Harrow and A. Hassidim and S. Lloyd",
   title        = "Quantum algorithm for linear systems of equations",
   journal      = "Phys. Rev. Lett.",
   volume       = "103",
   number       = "15",
   pages        = "150502",
   year         = "2009",
   month        = "Oct",
   doi          = "10.1103/PhysRevLett.103.150502"
}

@ARTICLE{Peruzzo2014,
   author       = "A. Peruzzo and J. McClean and P. Shadbolt and M.-H. Yung and X.-Q. Zhou and P. J. Love and A. Aspuru-Guzik and J. L. O'Brien",
   title        = "A variational eigenvalue solver on a photonic quantum processor",
   journal      = "Nat. Commun.",
   volume       = "5",
   pages        = "4213",
   year         = "2014",
   month        = "Jul",
   doi          = "10.1038/ncomms5213"
}

@ARTICLE{Gidney2021Stim,
   author       = "C. Gidney",
   title        = "{Stim}: A fast stabilizer circuit simulator",
   journal      = "Quantum",
   volume       = "5",
   pages        = "497",
   year         = "2021",
   month        = "Jul",
   doi          = "10.22331/q-2021-07-06-497"
}

@INPROCEEDINGS{KingmaBa2015,
   author       = "D. P. Kingma and J. Ba",
   title        = "{Adam}: A method for stochastic optimization",
   booktitle    = "Proc. Int. Conf. Learn. Representations (ICLR)",
   year         = "2015",
   month        = "May",
   pages        = "1--15",
   doi          = "10.48550/arXiv.1412.6980"
}

\end{document}